\documentclass[twocolumn]{aastex631}

\usepackage{amsmath, amssymb, amsfonts}
\usepackage{mathtools}
\usepackage{bm}
\usepackage{url}
\usepackage{cancel}
\usepackage{epstopdf}
\usepackage{subfigure}

\begin{document}

\title{Low-mass Active Galaxies in the SAMI Galaxy Survey with Spatially-resolved Spectroscopy}

\author{Stellan Bechtold}
\affil{Department of Physics, Montana State University, Bozeman, MT 59717, USA}

\author{Amy Reines}
\affil{Department of Physics, Montana State University, Bozeman, MT 59717, USA}

\begin{abstract}

The smallest supermassive black holes (BHs), which provide constraints on BH seeds, reside in low-mass galaxies. Here, we present a systematic analysis of 990 low-mass galaxies in the SAMI Galaxy Survey to identify emission from accreting BHs using integral field spectroscopy (IFS). Employing a novel automated scoring algorithm based on spatially-resolved narrow emission line diagnostics, we find signatures of active galactic nuclei (AGNs) in 41 galaxies, as well as an additional 46 (less secure) candidates. The galaxies have stellar masses between $10^{9.4} \lesssim M_\star/M_\odot \lesssim 10^{10}$ (down to $10^{8.5}$ including less secure candidates), redshifts $z \lesssim 0.06$, and morphologies ranging from early-type ellipticals to late-type spirals. Our AGN fraction of 4\% (9\% if we include the less secure candidates) is significantly higher than those reported by studies using single-fiber spectroscopy ($\lesssim 1-2\%$). Indeed, our additional analysis of single-fiber spectra of the objects in our sample demonstrates that many of our AGN candidates detected via IFS are missed. This work
highlights the advantages of IFS, especially the ability to capture extended or decentralized emission from accreting BHs.  

\end{abstract}

\section{Introduction} \label{sec:intro}

Almost all massive galaxies host supermassive black holes (BHs) at their centers \citep{kormendy_inward_1995,ferrarese_fundamental_2000,kormendy_coevolution_2013}. It is less clear how common such BHs are in low-mass galaxies ($M_\star \lesssim 10^{10} M_{\odot}$), yet determining the BH occupation fraction in low-mass galaxies is expected to help constrain the BH seeding mechanism(s) in the early Universe \citep{volonteri_evolution_2008,ricarte_observational_2018}. The origins of BH seeds remain poorly understood, with models divided into "light" seeds and "heavy" seeds, \citep[e.g., ][]{volonteri_formation_2010,woods_titans_2019}. Light BH seeds may form from the remnants of Population III stars \citep{bromm_first_2011}, while heavy BH seeds could arise from the direct collapse of massive gas clouds \citep{loeb_collapse_1994,begelman_formation_2006,latif_turbulent_2022} or runaway collisions in dense star clusters \citep{portegies_zwart_formation_2004,devecchi_formation_2009,miller_upper_2012}. 
In high-mass galaxies, information about BH seeding is often obscured by frequent mergers and subsequent growth. In contrast, low-mass galaxies experience fewer mergers and BH growth \citep{bellovary_first_2011}, making their BHs more representative of the initial seed population. Light seeding models are expected to result in a near universal BH occupation for both massive and low-mass galaxies, while heavy seeding models are expected to result in lower occupation fractions for low-mass galaxies \citep[e.g.,][]{volonteri_evolution_2008,ricarte_observational_2018}.

Determining the true BH occupation fraction in low-mass galaxies remains extremely challenging, as dormant low-mass BHs cannot currently be detected beyond a few Mpc. However, active galactic nuclei (AGNs) in low-mass galaxies can provide lower limits on the BH occupation fraction, although these constraints are wavelength-dependent and influenced by survey characteristics such as sensitivity and angular resolution. Considerable effort has therefore been devoted to identifying AGN in low-mass galaxies using a wide range of techniques across multiple wavelength regimes \citep[see reviews by][]{greene_intermediate-mass_2020, reines_hunting_2022}. These efforts have produced a growing number of measurements of the AGN fraction in low-mass systems \citep[e.g.,][]{reines_dwarf_2013, pardo_x-ray_2016, mezcua_intermediate-mass_2018, wylezalek_sdss-iv_2018, birchall_x-ray_2020, salehirad_hundreds_2022, bykov_srgerosita_2024, mezcua_manga_2024, pucha_tripling_2025}.

Among optical methods, the most widely used techniques for identifying AGN in low-mass galaxies include emission line diagnostic diagrams such as the BPT diagram \citep{baldwin_classification_1981} and the VO87 diagrams \citep{veilleux_spectral_1987} (see \S\ref{sec:AGN_diagnostics} for further details), as well as the detection of broad Balmer line emission. These approaches have been applied extensively to single-fiber spectroscopic survey data to determine the dominant ionization mechanisms in galaxy centers and to identify AGN candidates in large samples of low-mass galaxies \citep{reines_dwarf_2013, moran_black_2014, salehirad_hundreds_2022, pucha_tripling_2025}. The application of these techniques has led to a substantial increase in the number of known AGN in low-mass galaxies in recent years. However, commonly used optical diagnostic diagrams can be biased against faint and low-metallicity galaxies \citep{groves_emission-line_2006, cann_limitations_2019}, which are prevalent in the low-mass regime, while broad-line AGN selection can be contaminated by supernova emission \citep{reines_dwarf_2013, baldassare_multi-epoch_2016}.

Alternative identification methods are employed to help mitigate these limitations and provide multi-wavelength support of AGN candidates, including X-ray observations \citep{reines_candidate_2014, lemons_x-ray_2015, pardo_x-ray_2016, baldassare_x-ray_2017, chen_hard_2017, mezcua_intermediate-mass_2018, birchall_x-ray_2020, eberhard_new_2025}, radio observations \citep{reines_he210_2011, mezcua_radio_2019, reines_new_2020, davis_radio_2022, eberhard_dwarf_2025}, optical and infrared variability \citep{baldassare_identifying_2018, baldassare_search_2020, martinez-palomera_introducing_2020, secrest_low_2020, ward_variability-selected_2022}, and coronal emission lines \citep{cann_hunt_2018, molina_sample_2021, salehirad_hundreds_2022}.

X-ray and radio observations are particularly effective at detecting off-nuclear AGN emission and wandering BHs, which are predicted to be more common in low-mass galaxies \citep{mezcua_intermediate-mass_2018, reines_new_2020, bellovary_2021}. However, these observations are observationally expensive and therefore not well suited for constructing large AGN candidate samples. Recently, several studies have instead applied optical diagnostic techniques to spatially resolved data from integral field spectroscopy (IFS) to identify both nuclear and off-nuclear AGN emission \citep{johnston_beyond_2023}. These studies generally report higher AGN fractions in low-mass galaxies compared to single-fiber spectroscopic techniques \citep{wylezalek_sdss-iv_2018, mezcua_hidden_2020, mezcua_manga_2024}.

In this paper, we provide a novel AGN candidate classifying scheme specialized for IFS data from the Sydney-Australian-Astronomical-Observatory Multi-object Integral-Field Spectrograph Galaxy Survey \citep[SAMI;][]{croom_sami_2021} in the low-mass regime. We assign an AGN likelihood score based on the presence of significant AGN emission in a galaxy using standard optical emission line diagrams. We detail the SAMI survey, data products, and parent sample criteria in Section \ref{sec:data}. We provide details on the optical diagnostic diagrams, how we use them to determine our final AGN samples, and how the results compare to single-fiber techniques in Section \ref{sec:results_analysis}. In Section \ref{sec:discussion}, we discuss the interesting AGN emission distributions in the sample and how our results compare to past low-mass AGN studies. Section \ref{sec:conclusion} presents our conclusions.

\begin{figure*}[t!]
    \centering
    \includegraphics[width=0.9\textwidth]{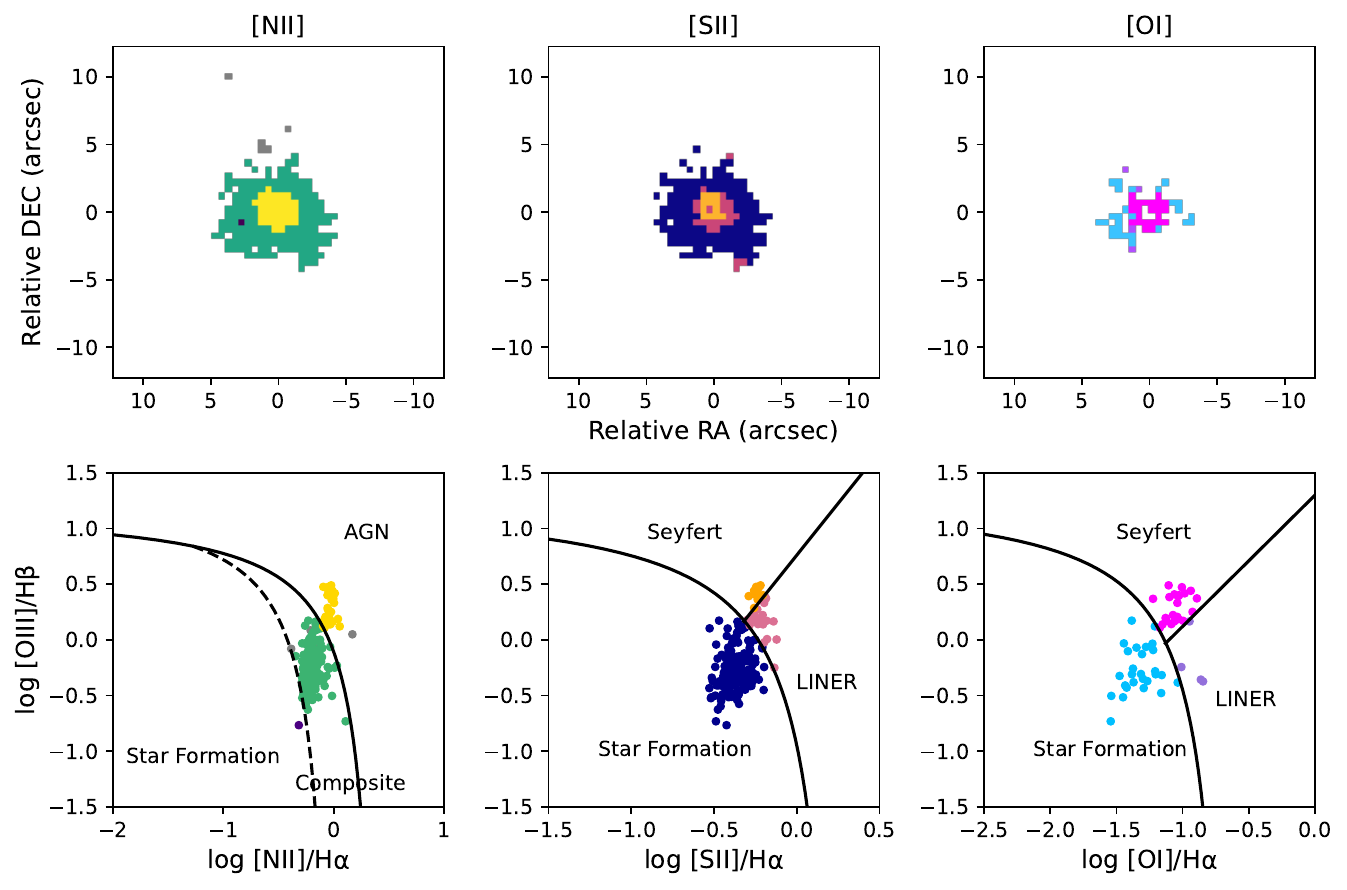}
    \caption{An example of spaxel maps (top row) and emission line diagnostic diagrams (bottom row) used in this work, as defined in \citet{kewley_host_2006},
    for SAMI galaxy CATID 287827. The top row shows the spatial distribution of spaxels classified by each of the three diagnostic diagrams:  
    [NII] diagram – star-forming (purple), Composite (green), and AGN (yellow);  
    [SII] diagram – star-forming (blue), LINER (red), and Seyfert/AGN (orange);  
    [OI] diagram – star-forming (light blue), LINER (light purple), and Seyfert/AGN (pink).
    Spurious AGN and Composite spaxels (see \S\ref{sec:selection_scheme}) are colored gray.
    White indicates omitted spaxels due to either a lack of data or high uncertainty in the flux measurements for that spaxel (see \S\ref{sec:parent_sample}).
    The coordinates in the diagnostic maps are relative to the galactic center as supplied by SAMI.
    The bottom row shows the same spaxels plotted on their respective diagnostic diagrams.}
    \label{fig:287827_bpt}
\end{figure*}

\section{Data} \label{sec:data}
\subsection{IFS from the SAMI Galaxy Survey} \label{sec:SAMI_survey}

We utilize the third and final data release (DR3) of the SAMI galaxy survey, which contains 3068 unique galaxies with redshifts ranging from $0.004 \leq z \leq 0.113$ and stellar masses in the range $7.5 \leq \log_{10}(M_*/M_{\odot}) \leq 11.6$ \citep{scott_sami_2018}. The stellar masses $(M_*/M_{\odot})$ are estimated using $i-$band magnitudes and $g - i$ colors \citep{taylor_galaxy_2011,bryant_sami_2015}. 
Most of the target galaxies in the SAMI survey were selected from the three equatorial regions (G09, G12, G15) of the Galaxy and Mass Assembly (GAMA) Survey, but a significant portion of the sample (888 of the 3068) are from eight previously unobserved galaxy clusters \citep{owers_sami_2017}. 

The SAMI survey collected optical integral field spectroscopy (IFS) data using the 3.9m Anglo-Australian Telescope. For more details on survey design and data products, see \citet{croom_sami_2021}. The SAMI instrument is a multi-object IFS system capable of observing up to 13 galaxies simultaneously. Each target galaxy is observed with a hexabundle of 61 fibers, providing a circular field of view with a radius of $15''$ per observation.

The spectral data for each galaxy are stored in two primary data cubes covering different wavelength ranges. The blue arm uses the 580V grating, covering 370--570 nm with a resolving power of $R = 1808$ (corresponding to $\sigma = 70.4\,\text{km}\,\text{s}^{-1}$), while the red arm uses the R1000 grating, covering 630--740 nm with $R = 4304$ ($\sigma = 29.6\,\text{km}\,\text{s}^{-1}$). Each data cube has a spatial pixel (spaxel) scale of $0.5''$, resulting in a $50 \times 50$ spatial grid and a total field of view of $25'' \times 25''$.

We used the SAMI DR3 value-added emission–line products generated with the LZIFU fitting pipeline \citep{ho_lzifu_2016} following the procedure described in \citet{green_sami_2018}, \citet{scott_sami_2018}, and the DR3 release paper \citep{croom_sami_2021}. In the SAMI fitting procedure, the stellar continuum in each spaxel is modeled using pPXF \citep{cappellari_parametric_2004,cappellari_improving_2017,cappellari_full_2023} with the MILES SSP library \citep{vazdekis_evolutionary_2010}, supplemented by young SSP templates from \citet{delgado_evolutionary_2005}.

Because the per-spaxel continuum S/N in the full-resolution cubes is often too low for a reliable direct pPXF fit, the continuum and emission lines are first fit simultaneously in Voronoi-binned spectra. The resulting template weights are then used as priors when refitting the individual spaxels in the full-resolution cube \citep{owers_sami_2019}. There is no S/N cut applied in the continuum fitting, in spaxels where the continuum S/N is zero or negative (due to sky-subtraction residuals) the continuum is set to zero, but LZIFU still attempts to fit any detectable emission lines. Spaxels lacking sufficient spectral coverage (more than 500 missing pixels in either the blue or red arm) are flagged and excluded from the emission line fitting process.

After continuum subtraction, LZIFU fits the emission lines H$\alpha$, H$\beta$, [OIII]$\lambda\lambda4959,5007$, [OII]$\lambda\lambda3726,3729$, [OI]$\lambda6300$, [NII]$\lambda\lambda6548,6583$, and [SII]$\lambda\lambda6716,6731$ simultaneously using one, two, or three Gaussian components. The recommended number of components is determined using the LZCOMP neural network \citep{hampton_using_2017}.

Although no explicit continuum S/N cut is applied in the SAMI pipeline, spaxels are required to meet minimum S/N thresholds for all emission lines entering the diagnostic diagrams in our selection scheme (see \S\ref{sec:selection_scheme}). In the SAMI pipeline, emission line flux uncertainties are derived from the full spectral fit, and uncertainties in the stellar continuum modeling should propagate into the emission line errors. As a result, spaxels with poorly constrained continua are expected to have larger emission line uncertainties and are therefore less likely to satisfy the emission line S/N requirements. Consequently, spaxels with very low continuum S/N are unlikely to significantly influence our results.

For this study we use $50\times50$ flux maps for the emission lines H$\alpha$, H$\beta$, [OIII]$\lambda5007$, [OI]$\lambda6300$, [NII]$\lambda6583$, [SII]$\lambda6716$, and [SII]$\lambda6731$, adopting the single-Gaussian component fits provided in the DR3 products for consistency across the full sample. Additional details on the SAMI emission-line fitting procedures can be found in \citet{green_sami_2018}, \citet{scott_sami_2018}, and \citet{croom_sami_2021}.

\subsection{Single-Fiber Spectroscopy} \label{sec:single_fiber_spectroscopy}

Most target galaxies in the SAMI Galaxy Survey, excluding the cluster galaxies, have associated single-fiber spectra from previous surveys such as GAMA \citep{driver_gama_2009,driver_galaxy_2011}, Sloan Digital Sky Survey \citep[SDSS;][]{york_sloan_2000}, 2dF Galaxy Redshift Survey \citep[2dFGRS;][]{colless_2df_2001, madgwick_2df_2002}, and 6dF Galaxy Survey \citep[6dFGS;][]{jones_6df_2004,jones_6df_2009} (see \citealt{baldry_galaxy_2018} for details). After identifying AGN candidates in low-mass galaxies using IFS from the SAMI Galaxy Survey (\S \ref{sec:selection_scheme}), we analyze the single-fiber spectra of these objects for comparison using the procedure described in Section \ref{sec:single_fiber}.

\begin{figure*}[t!]
    \centering
    \includegraphics[width=\textwidth]{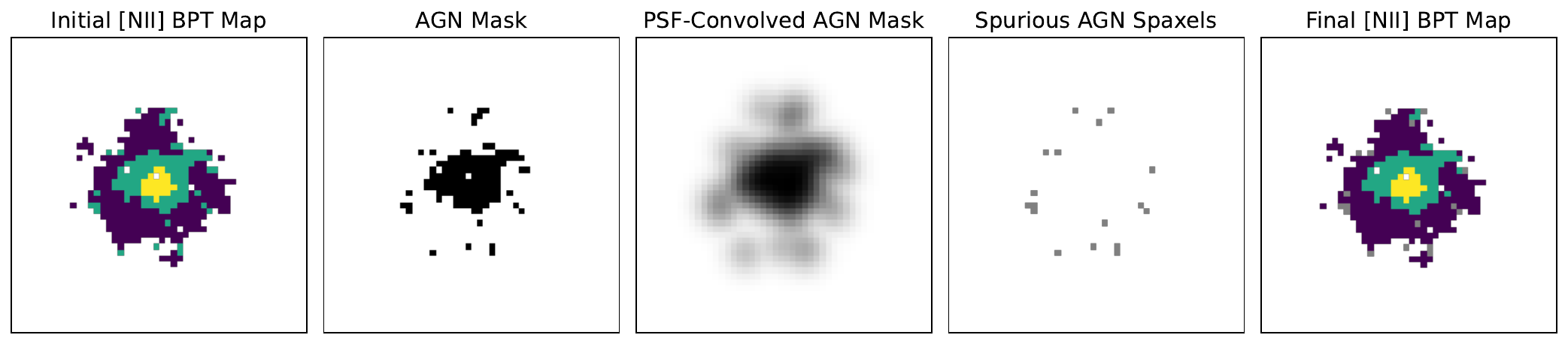}
    \caption{Example of the AGN spaxel clustering analysis for galaxy CATID 382563. Left to right: (1) Initial [NII] diagnostic map showing AGN (yellow), Composite (green), and star-forming (purple) spaxels (see Figure \ref{fig:287827_bpt} for color scheme). (2) Boolean AGN mask in which AGN and Composite spaxels are assigned a value of one and all other spaxels zero. (3) PSF-convolved AGN mask using a 2D Gaussian kernel ($\sigma = 1.7$ spaxels, corresponding to $2''$ FWHM). (4) Spaxels with PSF-weighted values below the adopted threshold of 0.2, shown in gray and considered spurious (see \S\ref{sec:selection_scheme}). (5) Final [NII] BPT map with spurious AGN classifications re-labeled in gray.}
    \label{fig:psf_example}
\end{figure*}

\subsection{Parent Sample of Low-Mass Galaxies} \label{sec:parent_sample}

The primary goal of this project is to identify AGNs in low-mass galaxies. We now describe how we construct a parent sample of galaxies based on stellar mass and data quality.
First, we impose a stellar mass threshold of $M_* \leq 10^{10} \, M_{\odot}$, yielding an initial sample of 1201 galaxies. This limit extends beyond the conventional dwarf galaxy cutoff of $3 \times 10^9 \, M_{\odot}$ used in previous studies such as \citet{reines_dwarf_2013}, but is consistent with the threshold adopted in \citet{salehirad_hundreds_2022}. Using this higher mass ceiling facilitates a comparison between our AGN detections in SAMI and that of \citet{salehirad_hundreds_2022}, who used single-fiber spectra from GAMA.

Second, we apply a data quality requirement to ensure that emission line measurements are reliable. For a spaxel to be included in any diagnostic diagram, it must meet a minimum signal-to-noise ratio of S/N $\geq$ 3 for all four emission lines required by that diagram. We define the S/N of each emission line as the line flux divided by its $1\sigma$ uncertainty, as provided by the SAMI data products (see \S\ref{sec:SAMI_survey}). Since the required lines differ between diagnostics, a spaxel may qualify in one diagram but not in another. Spaxel classifications may also differ across diagrams.

To ensure sufficient diagnostic coverage, we additionally require that each galaxy has at least 5 spaxels that meet S/N requirements in the [NII] diagram. This threshold removes low-quality observations and ensures that each galaxy contributes meaningful spatial information to our analysis.

We demand sufficient spaxels in the [NII] diagram as opposed to the [SII] or [OI] diagrams because the [NII] diagram is the primary diagnostic used in our AGN candidate selection scheme (see \S\ref{sec:selection_scheme} for details). 

To summarize our requirements on the parent sample of low-mass galaxies:

\begin{enumerate}
    \item Stellar mass cut: $M_* \leq 10^{10} \, M_{\odot}$
\end{enumerate}

\begin{enumerate}
    \setcounter{enumi}{1}
    \item Data quality cut: A galaxy must have at least 5 spaxels in the [NII] diagram, where each spaxel much have S/N $\geq$ 3 for all relevant emission lines in the diagnostic.
\end{enumerate}

\noindent
Applying the $M_*$ cutoff removes galaxies with a reported stellar mass greater than $10^{10} \, M_{\odot}$ as well as those with no reported stellar mass, reducing the initial SAMI sample from 3068 to 1201 galaxies. The subsequent spaxel quality and minimum valid spaxel requirements further reduce this to a final parent sample of 990 galaxies. These 990 galaxies serve as the input to our specialized low-mass AGN selection scheme described below. \\

All data used in this paper can be accessed online from Australian Astronomical Optics Data Central: \url{https:/datacentral.org.au/}.

\section{Analysis and Results} \label{sec:results_analysis}

\subsection{Emission Line Diagnostics} \label{sec:AGN_diagnostics}

The most widely used tools for identifying AGN in optical spectroscopic surveys are the BPT \citep{baldwin_classification_1981} and VO87 \citep{veilleux_spectral_1987} diagnostic diagrams, which distinguish between gas ionization sources based on emission line flux ratios. The three main diagnostics used are: 
\begin{itemize}
    \item $[OIII]\lambda5007/H\beta \text{ vs. }[NII]\lambda6583/H\alpha$ \\ $\text{ ([NII] diagram)}$,
    \item $[OIII]\lambda5007/H\beta \text{ vs. }[SII]\lambda6716,6731/H\alpha$ \\ $\text{ ([SII] diagram)}$,
    \item $[OIII]\lambda5007/H\beta \text{ vs. }[OI]\lambda6300/H\alpha$ \\ $\text{ ([OI] diagram)}$.
\end{itemize}

Following the classification scheme presented in \citet{kewley_host_2006}, the [NII] diagram separates galaxies into three regions: star forming (SF), Composite, and AGN. Galaxies above the theoretical SF boundary \citep{kewley_theoretical_2001} require additional ionization sources (i.e., a dominant AGN), with the Composite region reflecting a mix of SF and AGN emission \citep{kauffmann_host_2003}. The [SII] and [OI] diagrams provide further separation of AGN into Seyfert and Low-Ionization Nuclear Emission-Line Region (LINER) subclasses. While LINER emission can be associated with AGN activity, it may also arise from non-AGN sources such as stellar radiation, galactic winds, or supernova-driven shocks \citep{veilleux_spectral_1987, molina_shocking_2018}.

In traditional single-fiber surveys, each galaxy is represented by a single point in the diagnostic diagrams, derived from a central spectrum. In contrast, integral field spectroscopy (IFS) enables spatially resolved classification, where each galaxy contributes many spaxels that can be independently placed on the diagnostic diagrams. This approach allows us to trace AGN and SF regions within galaxies and study their spatial distribution (see \S\ref{sec:SAMI_survey}). Examples of these resolved diagnostic diagrams are shown in the bottom row of Figure \ref{fig:287827_bpt}.

Among the three diagnostic diagrams, the [NII] diagram is the most widely used, including studies that look for AGN in the low-mass regime with single-fiber surveys \citep{reines_dwarf_2013,salehirad_hundreds_2022, pucha_tripling_2025} and IFS data \citep{mezcua_manga_2024}. Using the [NII] diagram as our primary diagnostic enables better comparison with previous works. However, [NII] classifications are sensitive to metallicity, which poses a challenge in low-mass, metal-poor galaxies. The [SII] and [OI] diagrams are less sensitive to metallicity and can serve as useful complementary diagnostics for identifying AGN in this regime \citep{polimera_resolve_2022}.

It is also worth noting that low-mass galaxies are expected to host AGNs with lower luminosities given their smaller BHs. The AGN emission in these systems can also be more easily diluted or obscured by star formation \citep{groves_emission-line_2006, stasinska_semi-empirical_2006, cann_limitations_2019}. Together, these introduce a detection bias in emission line ratio diagnostics from a flux-limited survey toward high-luminosity, high-accretion AGNs \citep{kewley_understanding_2019}.

\subsection{AGN Spaxel Clustering} \label{sec:spaxel_clustering}

For each of the 990 low-mass galaxies in our parent sample (\S\ref{sec:parent_sample}), we classify each spaxel using the three emission line diagnostics discussed in \S\ref{sec:AGN_diagnostics}. We then map the spatial distribution of emission-line classifications by assigning each spaxel a color according to its diagnostic class (Figure \ref{fig:287827_bpt}). In many galaxies, these maps reveal isolated AGN-classified spaxels. Since individual spaxels ($0.5'' \times 0.5''$) are substantially smaller than the typical SAMI seeing PSF with a FWHM $\sim 2''$ \citep[e.g.][]{croom_sami_2021, zovaro_sami_2024}, such isolated spaxels may be spurious and cause misleading results.

To reduce spurious AGN emission in our sample, we account for the seeing PSF and consider the grouping of AGN spaxels as follows. For each diagnostic map, we construct a boolean AGN mask in which AGN classified spaxels (and Composite spaxels in the [NII] diagram) are assigned a value of one and all other spaxels are assigned zero. We convolve this mask with a two-dimensional normalized Gaussian PSF with $\sigma = 1.7$ spaxels, corresponding to a FWHM of $\sim 2''$ at the SAMI spaxel scale ($0.5''$/spaxel). A single isolated AGN spaxel produces a peak PSF-weighted value of $\sim 0.055$. To suppress such isolated classifications while remaining sensitive to compact AGN emission, we adopt a PSF weighted threshold of 0.2. This threshold effectively requires multiple neighboring AGN classified spaxels within a single PSF footprint, enforcing spatial coherence at the instrumental resolution. Each diagnostic map is treated independently in this procedure. AGN spaxels failing to meet the PSF threshold are labeled in gray on the diagnostic diagrams and maps. A step-by-step example of this clustering analysis for the [NII] diagnostic map of galaxy CATID 382563 is shown in Figure \ref{fig:psf_example}.

Only AGN classified spaxels meeting the PSF weighted threshold contribute to the final selection scheme and scoring. Spaxels classified as SF or LINER are not promoted to AGN classification based on proximity to AGN spaxels.

\begin{figure}[t!]
    \centering
    \includegraphics[width=3.25in]{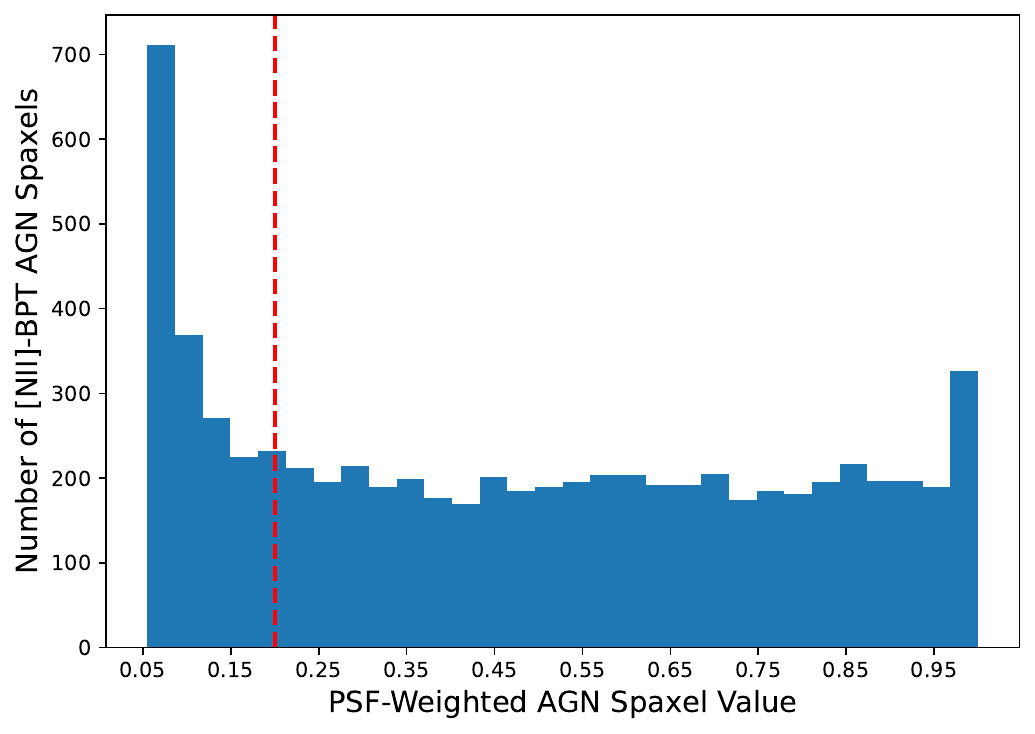}
    \caption{Distribution of PSF-Weighted AGN Score for all [NII]-BPT AGN or Composite classified spaxels in the parent sample. Red dotted line indicates our adopted PSF value threshold of 0.2.}
    \label{fig:psf_value_hist}
\end{figure}

To guide our choice of PSF weighted AGN threshold, we examined the distribution of PSF values across all [NII]-BPT AGN/Composite spaxels. As shown in Figure \ref{fig:psf_value_hist}, the distribution shows a steep decline at low values, followed by a relatively flat plateau beginning at $\sim0.2$ and extending to $\sim1$, with a sharp rise at the highest values corresponding to strongly AGN-dominated regions. We interpret the rapid decline at low values as reflecting spaxels that are mostly isolated without substantial contributions from neighboring spaxels, while the plateau represents a population of spatially coherent AGN emission. Motivated by this transition, we adopt a threshold of 0.2, which lies at the boundary between these regimes and provides a conservative balance between suppressing spurious spaxels and retaining physically meaningful AGN structure.

\subsection{AGN Selection Scheme and Final Sample} \label{sec:selection_scheme}

While the field of AGN classification via IFS data and optical diagnostics is still relatively sparse, there are a small number of papers that have developed various classification schemes that inform our method. Diagnostic diagrams have been employed to identify AGN candidates across all stellar masses using IFS data from both the MaNGA survey \citep{wylezalek_sdss-iv_2018} and SAMI \citep{johnston_beyond_2023}. In addition, there has been previous work that has specialized in identifying AGNs specifically in the low-mass regime using MaNGA IFS data and optical diagrams \citep{mezcua_manga_2024}. A common theme among these studies is an established threshold for what constitutes sufficient AGN emission to identify a candidate. The threshold could be a minimum number of AGN or Composite dominated spaxels present in the galaxy (e.g., 10 as in \citealt{johnston_beyond_2023}), or it could be a minimum portion of the spaxels in the galaxy that lie in the AGN or Composite regions of the diagnostic diagrams (e.g., 5$\%$ as in \citealt{mezcua_manga_2024}). The three mentioned papers also require that there must be an AGN signal in more than one of the diagrams to allow a galaxy to achieve candidate status.

Again, while our method is informed by studies such as those mentioned above, we tailor our selection scheme based on the SAMI data for galaxies in the low-mass regime. Given the biases mentioned at the end of \S\ref{sec:AGN_diagnostics}, we have tuned our scheme to accommodate detection of low-level AGN emission in low-mass galaxies with sometimes limiting information available for optical diagnostics.

After removing spurious AGN spaxels (\S\ref{sec:spaxel_clustering}), we create an AGN host likelihood rating, or "score", based on the three diagnostic diagrams. Our scoring system is informed by previous work on AGN classification using IFS data, but unique in terms of the details as this is the first search specifically for AGNs in low-mass galaxies in the SAMI Galaxy Survey.

In our selection scheme, galaxies begin with a score of zero and accumulate likelihood points based on evidence of AGN emission in the three diagnostic diagrams. We identify AGN presence using a minimum number of AGN spaxels rather than a proportion, as the latter can be misleading for low-mass galaxies. For example, AGN signatures may be concentrated in small regions but outnumbered by SF spaxels since low-mass galaxies are star-forming on the whole.

\begin{figure*}[t!]
    \centering
    \includegraphics[width=0.88\textwidth]{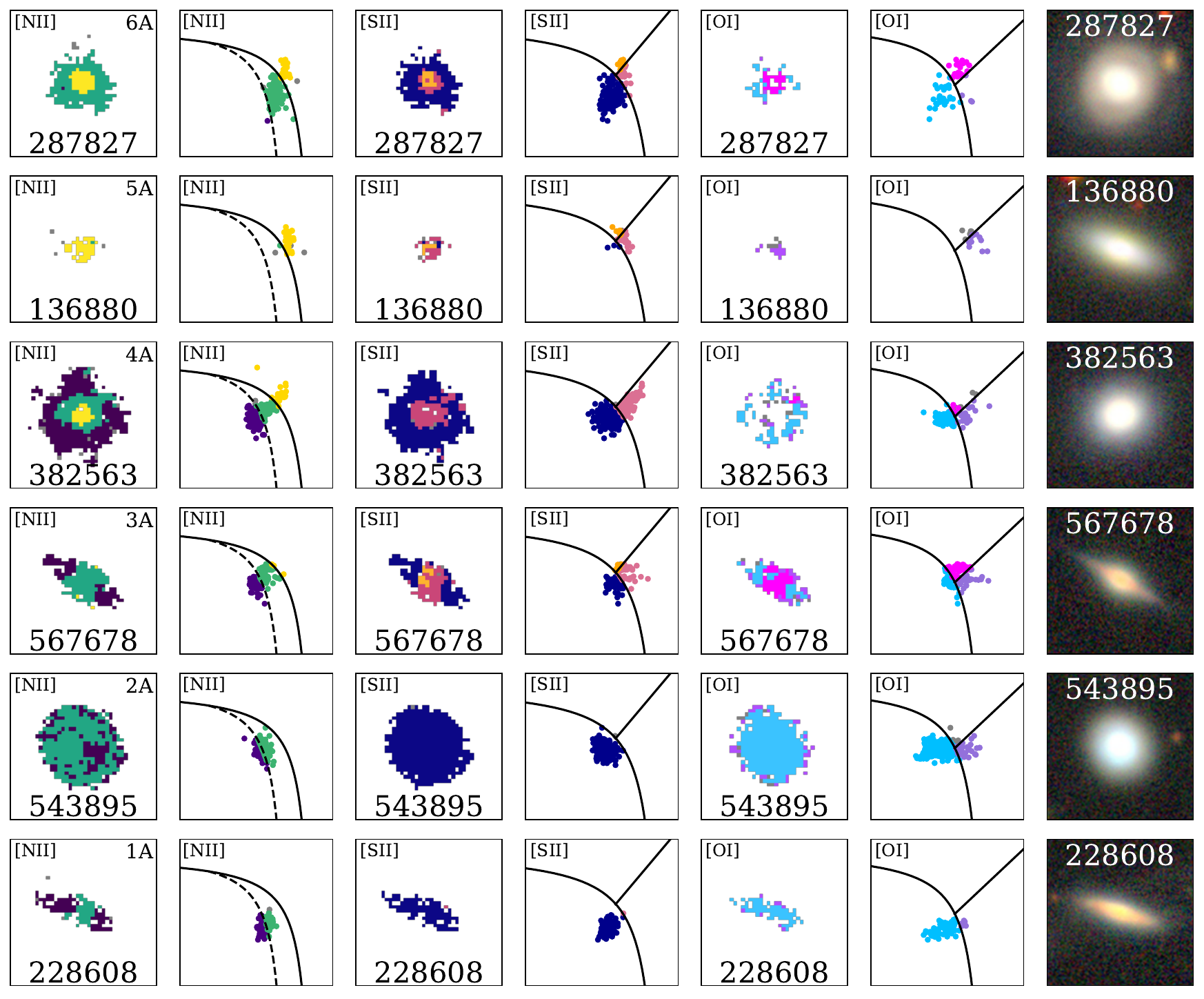}
    \caption{
    Example galaxies in Sample A. We show a representative galaxy for each score value of 1 to 6, with a larger score indicating more evidence for an AGN. The score is shown in the top right corner of the first column for each row, with the accompanying letter indicating the galaxies are in Sample A. The first six columns show the [NII], [SII], and [OI] classification spatial maps interspersed with the corresponding diagnostic diagrams. The color code for all maps and diagrams follows Figure \ref{fig:287827_bpt}. The final column shows the $grz$-band $25'' \times 25''$ optical cutout from the DESI Legacy Imaging Survey.}
    \label{fig:score_sample}
\end{figure*}

The [NII] diagnostic serves as the primary indicator of AGN activity in our selection scheme, while the [SII] and [OI] diagrams provide secondary supporting evidence. Galaxies accumulate points based on the presence of AGN emission in these diagrams, subject to a minimum requirement of five spaxels in the relevant classification region. Only galaxies that exhibit AGN or Composite emission in the [NII] diagram are eligible for inclusion in the final sample and galaxies lacking such emission are classified as star-forming (SF) and excluded regardless of their classifications in the other diagrams. Spaxels classified as LINERs are not counted toward AGN spaxel totals in any diagnostic, in order to avoid ambiguous cases (see \S\ref{sec:AGN_diagnostics}).

The scoring procedure is defined as follows:

\begin{enumerate}
\item Each galaxy begins with a score of 0.
\item If the [NII] diagram contains at least 5 AGN classified spaxels, the galaxy receives +4 points.
\item If fewer than 5 AGN spaxels are present but at least 5 AGN or Composite spaxels are present in the [NII] diagram, the galaxy receives +1 point.
\item Galaxies with a score of 0 after evaluation of the [NII] diagram are classified as star-forming and removed from the AGN candidate sample.
\item For the remaining galaxies, the presence of at least 5 Seyfert classified spaxels in the [SII] or [OI] diagrams contributes +1 point per diagram.
\end{enumerate}

\begin{deluxetable}{ccc}
\tabletypesize{\footnotesize}
\tablecolumns{2} 
\tablecaption{\label{tab:sample_stats}Sample Statistics} 
\tablehead{\colhead{AGN Score} & \colhead{\hspace{.75cm}Sample A}\hspace{.5cm} & \colhead{\hspace{.5cm}Sample B}\hspace{.75cm}} 
\startdata
1 & 21 & 36 \\
2 & 4 & 7 \\
3 & 1 & 1 \\
4 & 7 & 0 \\
5 & 4 & 0 \\
6 & 4 & 2 \\
\hline
Total & 41 (4.1$\%$) & 46 (4.6$\%$) \\
\enddata 
\tablecomments{Galaxy counts for each AGN score across samples A and B. Final percentage shows portion of parent sample of 990 galaxies.}
\end{deluxetable}

Scores can range from 0 to 6. A score of zero indicates little to no evidence for AGN presence in the galaxy, and higher scores indicate more evidence of AGN emission with a score of six indicating substantial evidence for AGN presence. The scoring system was intentionally designed so that Composite and AGN galaxies as defined by the [NII] diagram can be clearly separated into score groups [1,2,3] and [4,5,6], respectively. A galaxy that is classified as [NII] Composite cannot receive as many points as an [NII] AGN galaxy, even with additional points from the [SII] and [OI] diagrams. There are 903 galaxies that received a zero and are excluded from the final AGN sample. The remaining 87 galaxies that received a non-zero score make up the final AGN sample.
Therefore, our final AGN sample consists of approximately $9\%$ of the parent sample (87/990). 
In Figure \ref{fig:score_sample}, we present an example galaxy for each scoring bin from 1 to 6 in our final sample to illustrate the scoring system and provide a clearer visual understanding of the classification process.

\begin{figure}
    \centering
    \includegraphics[width=\linewidth]{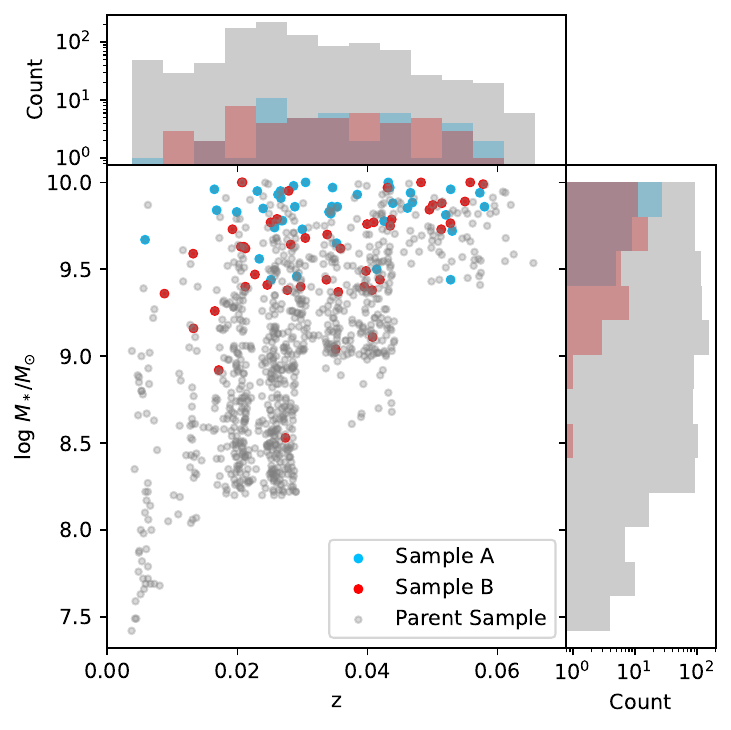}
    \caption{\textit{Center panel}: Galaxy stellar mass versus redshift. Sample A galaxies are shown in blue, Sample B galaxies are in red, and our parent sample of 990 low-mass galaxies is shown in gray. \textit{Top panel}: Redshift (z) histograms of the three samples. \textit{Right panel}: $\log M_*/M_{\odot}$ histograms of the three samples.}
    \label{fig:mass_z_joint}
\end{figure}

We lower the minimum AGN spaxel requirement from the 10 spaxels adopted in previous studies (e.g., \citealt{johnston_beyond_2023}) to 5 spaxels in order to reduce bias against less luminous AGN in low-mass galaxies. To quantify the impact of this choice, we selected galaxies from the parent sample using both 5 and 10 spaxel thresholds based on the [NII] diagnostic alone. The 5 spaxel threshold yields a sample of 87 galaxies, whereas the 10 spaxels threshold reduces the sample to 60 galaxies, a decrease of $\sim31\%$. The difference in the median stellar mass between the 27 excluded galaxies (median $\log M_*/M_{\odot} = 9.62$) and the 60 galaxies retained under both criteria (median $\log M_*/M_{\odot} = 9.79$) is modest. However, a Kolmogorov--Smirnov test indicates that the two groups of stellar masses are sampled from statistically distinct populations ($p \approx 0.017$). This demonstrates that a 10 spaxel threshold preferentially excludes lower-mass systems. We therefore adopt a 5 spaxel requirement, supplemented by spatial and PSF based validation, to mitigate this bias while still suppressing spurious AGN classifications.

\subsubsection{Uncertainty Test and Defining Samples A and B} \label{sec:uncertainty_test}

Visual inspection of this AGN sample reveals that a significant number of galaxies share common characteristics. 
High-scoring galaxies typically show AGN emission in the central region of the galaxy, with a few exceptions. CATIDs 287827, 136880, 382563 and in Figure \ref{fig:score_sample} (top three rows with scores of 6, 5 and 4, respectively) are examples of high scoring galaxies with central AGN emission. We also see in these galaxies that the AGN spaxels are surrounded by Composite spaxels in many cases.
Galaxies with lower scores display a larger variety in spatial distribution of AGN emission compared to high scoring galaxies. There are galaxies that show only Composite emission from the central region, much like AGN galaxies (e.g., CATIDs 567678 and 228608 in the fourth and sixth rows of Figure \ref{fig:score_sample}, respectively). However, some galaxies contain Composite emission across the vast majority of the galaxy (e.g., CATID 543895 in the fifth row of Figure \ref{fig:score_sample} with a score of 2.)
There is also a large portion of the low scoring AGN sample where most of the Composite emission is decentralized, appearing away from the galactic center. We will discuss this in more detail in \S\ref{sec:decentralized_agn}.

We further divide our final AGN sample into Sample A and Sample B, by considering how uncertainties in the emission line fluxes impact AGN classification. We observe that the Composite spaxels lie close to the SF-Composite demarcation line in the [NII] diagram for many of the galaxies in our sample. Given the uncertainties in the emission line fluxes provided by the data processing pipeline (see \S\ref{sec:SAMI_survey}), it is possible that these Composite spaxels could shift into the SF region when the uncertainties are taken into account. To address this, we reprocess each of the 87 galaxies by incorporating the uncertainties in the positions of spaxels on the diagram. In this analysis, we simulate a "worst-case scenario" by shifting the position of each spaxel down and to the left on the [NII] diagram, toward the SF region, by their respective $2\sigma$ uncertainties along each axis. If a galaxy satisfies the selection criteria and remains in the AGN sample after this shift, it is assigned to Sample A. If a galaxy is filtered out by this process, but was included in the original 87 galaxy AGN sample, it is placed in Sample B.

Following this "worst-case scenario" analysis, Sample A contains 41 galaxies, while Sample B contains 46 galaxies. Notably, most galaxies in Sample B score in the 1–3 point range, meaning they show weaker evidence of AGN emission in the [NII] diagram and rely heavily on Composite emission for classification, except for 2 galaxies with scores of 6, which we discuss more in \S\ref{sec:decentralized_agn}. The breakdown of scores for Samples A and B is shown in Table \ref{tab:sample_stats}.

Diagnostic diagrams and maps for all galaxies in Sample A are available in Appendix \ref{sec:bpt_maps}. Additionally, we provide $grz$-band cutouts of each galaxy from the DESI Legacy Imaging Survey SkyViewer \citep{dey_overview_2019,xu_desi_2023}.

\begin{figure}
    \centering
    \includegraphics[width=\linewidth]{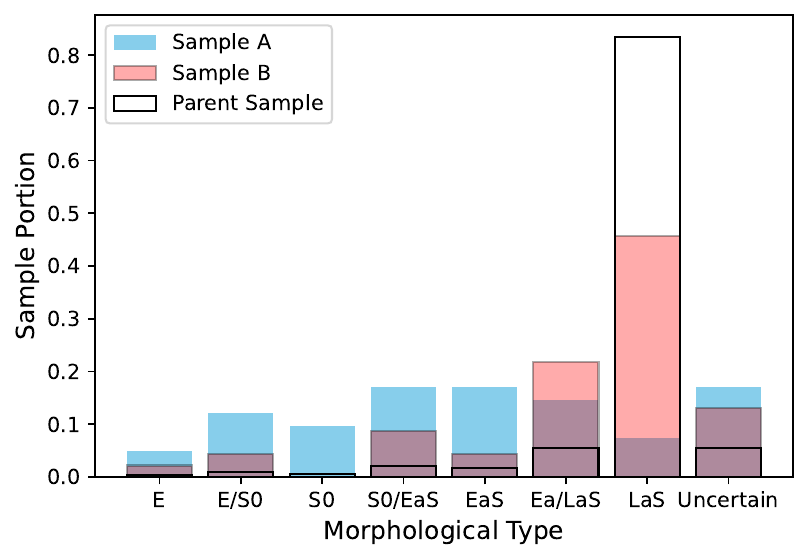}
    \caption{Histogram of morphological types for Samples A, B and the parent sample. Label definitions: E - elliptical, S0 - lenticular, EaS - early spiral, LaS - late spiral, Uncertain - subjective/no agreement.}
    \label{fig:morph_hist}
\end{figure}

\begin{figure*}[t!]
    \centering
    \includegraphics[width=\textwidth]{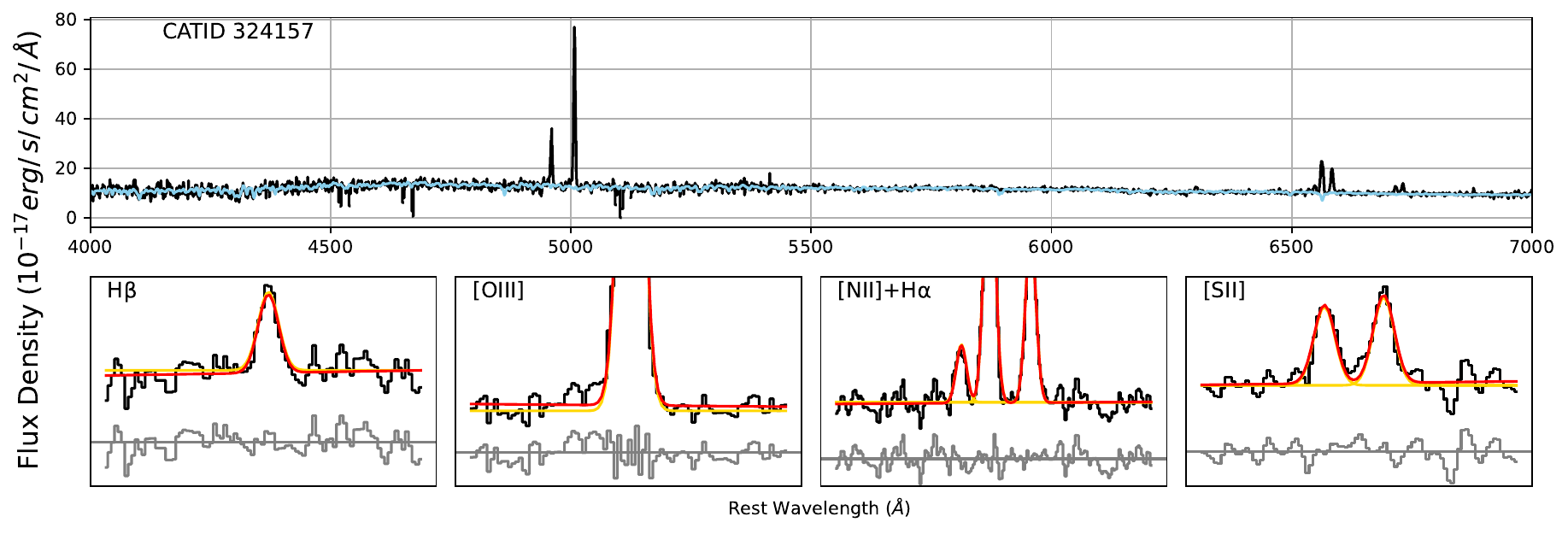}
    \caption{Example of our single-fiber spectral fitting for SAMI galaxy CATID 324157, which is included in Sample A with a score of 6. This spectrum is from the GAMA survey. The top panel shows the rest frame spectrum in black and the best-fit stellar continuum and absorption line model in light blue. The bottom four panels show sections of the spectrum used to fit the relevant emission lines. The spectrum is shown in black, Gaussian models of emission lines are shown in yellow, and final best-fit models for each section (Gaussian component(s) + linear component) are shown in red. Residuals are plotted in gray below each emission line region.}
    \label{fig:spec_grid}
\end{figure*}

\subsection{Host Galaxies}

In this section, we investigate the properties of galaxies with evidence of AGN emission in Samples A and B, and compare them to our parent sample of low-mass galaxies. The right panel of Figure \ref{fig:mass_z_joint} shows histograms of the stellar mass distribution for the AGN sample and the parent sample. Sample A spans a mass range of $\log (M_*/M_{\odot}) \approx 9.4$ to $10.0$, while the Sample B extends down to $\log (M_*/M_{\odot}) = 8.5$, and the parent sample extends down to $\log (M_*/M_{\odot}) = 7.4$. The lower limit of $\log (M_*/M_{\odot}) = 8.5$ is comparable to the minimum masses in prior dwarf AGN studies, such as \citet{reines_dwarf_2013}, which also found AGN candidates with the [NII] diagram down to $\log (M_*/M_{\odot}) \approx 8.5$. Based on the BH mass - stellar mass relation for local AGNs \citep{reines_relations_2015}, we estimate BH masses in the range of $\log (M_{\text{BH}}/M_{\odot}) \approx 4.8$ to $6.4$.

We note that the frequency of AGN candidates generally increases with higher mass values, especially in Sample A, up to our stellar mass limit of $10^{10} M_{\odot}$. This overall trend has also been seen in previous AGN studies \citep{kauffmann_host_2003, reines_dwarf_2013, baldassare_identifying_2018, salehirad_hundreds_2022, mezcua_manga_2024}.  Higher-mass galaxies tend to host more massive black holes, which, for a given Eddington ratio, are more luminous and thus easier to detect than their lower-mass counterparts. This introduces an observational bias that likely suppresses the observed AGN fraction at lower stellar masses.

The top panel of Figure \ref{fig:mass_z_joint} shows the redshift distributions of the parent sample, as well as Samples A and B. The three samples span a similar redshift range ($z \lesssim 0.065$), and the shapes of their distributions are broadly comparable. This suggests that any observed differences in AGN fraction between the samples are not primarily driven by redshift, as expected given the limited range probed by this study.

The center panel of Figure \ref{fig:mass_z_joint} shows the galaxy distribution in redshift–stellar mass space. As expected in a flux-limited survey, lower-mass galaxies are typically found at lower redshifts, where they are bright enough to meet survey S/N thresholds. Conversely, at higher redshifts the sample is dominated by more massive and luminous galaxies. The apparent decline in AGN candidates at the lowest redshifts likely reflects the fact that these systems are also the least massive and faintest, making their AGN emission more difficult to detect, and that the intrinsically rarer, more luminous AGNs are unlikely to be found within such a small volume. Whether this drop reflects a true decline in AGN fraction for low-mass systems or simply limited sensitivity remains uncertain.
We also note that the redshift–mass structure of the parent sample primarily reflects SAMI’s original target selection strategy, which was based on redshift and a proxy for stellar mass. This results in the step-like features visible in Figure \ref{fig:mass_z_joint}. For more on SAMI target selection, see \citet{bryant_sami_2015} and \citet{croom_sami_2021}.

The SAMI survey also supplies morphological types for each galaxy \citep{cortese_sami_2016}. Several members of the SAMI team visually classified each galaxy based off RGB combined color images from SDSS DR9 or VST ATLAS surveys \citep{croom_sami_2021}. The distribution of morphological types of Sample A, Sample B, and the parent sample are shown in Figure \ref{fig:morph_hist}. 
The galaxies in Sample A span the full range of morphological types, from early-type ellipticals to late-type spirals. A diversity in morphologies was also noted by \citet{kimbrell_diverse_2021} based on high-resolution observations of the \citet{reines_dwarf_2013} sample of AGNs in dwarf galaxies.
Sample B and the parent sample are dominated by late-type spiral galaxies. 
There are 6 galaxies in Sample A and 5 in Sample B that were considered too subjective or uncertain by the SAMI team to classify the morphology and are labeled "Uncertain" in the histogram.

\subsection{Comparison to Single-Fiber Spectra} \label{sec:single_fiber}

In this section, we aim to assess the effectiveness of IFS over single-fiber spectra in identifying AGNs in low-mass galaxies.
We retrieved single-fiber spectra for all available galaxies from the SAMI online data repository, measure emission line fluxes, and place objects on the [NII] diagram. Of the 87 galaxies across Samples A and B, 11 did not have available single-fiber spectra. These 11 galaxies are all from novel galaxy cluster observations conducted by the SAMI survey. The remaining 76 galaxies have available single-fiber spectra, and some galaxies have multiple spectra from one or more surveys. The spectra available in the SAMI database from the GAMA regions come from a suite of surveys: SDSS DR8, 2dFGRS, 6dFGS and GAMA (see \citet{baldry_galaxy_2018} for details about available spectra in GAMA regions). In these scenarios with multiple available spectra, we select the spectra labeled "best" by the online repository spectra for that galaxy as provided.

To obtain emission line flux measurements for $H\alpha$, $H\beta$, [NII], and [OIII] as required for the [NII] diagram, we follow a process similar to that described in \citet{reines_dwarf_2013} and \citet{salehirad_hundreds_2022}. The first step is to fit and remove the continuum and absorption lines in the spectra. We use the publicly available penalized pixel fitting code pPXF \citep{cappellari_full_2023} in which stellar continua are created using a linear combination stellar population synthesis model templates using the Flexible Stellar Population Synthesis package (fsps v3.2; \citealt{conroy_propagation_2009,conroy_propagation_2010}). After subtracting the best-fit stellar continuum and absorption line model, we fit emission lines with Gaussian profiles using the Python package LMFIT \citep{newville_lmfit_2014}. For each step in the fitting process described below, we select a chunk of spectrum around the line(s) of interest and include a linear component to account for any residual slope in the stellar continuum.

\begin{figure}
    \centering
    \includegraphics[width=\linewidth]{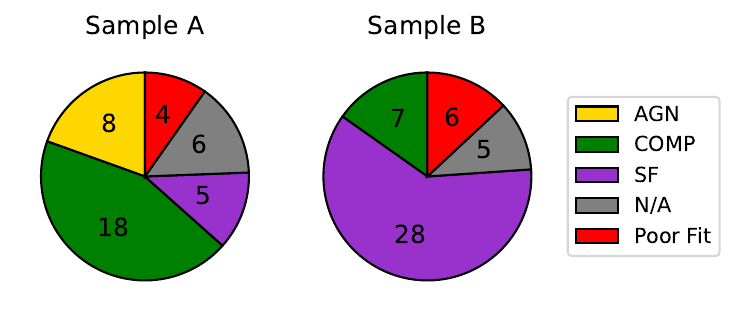}
    \caption{Pie charts of the [NII] diagram single-fiber classifications for galaxies in Samples A and B. In addition to the three typical classifications (SF, COMP, AGN), "N/A" refers to galaxies that had no available spectra, and "poor fit" is assigned to galaxies with low S/N in at least one of the relevant emission lines.} 
    \label{fig:1fib_piechart}
\end{figure}

\begin{figure*}
    \centering
    \includegraphics[width=\textwidth]{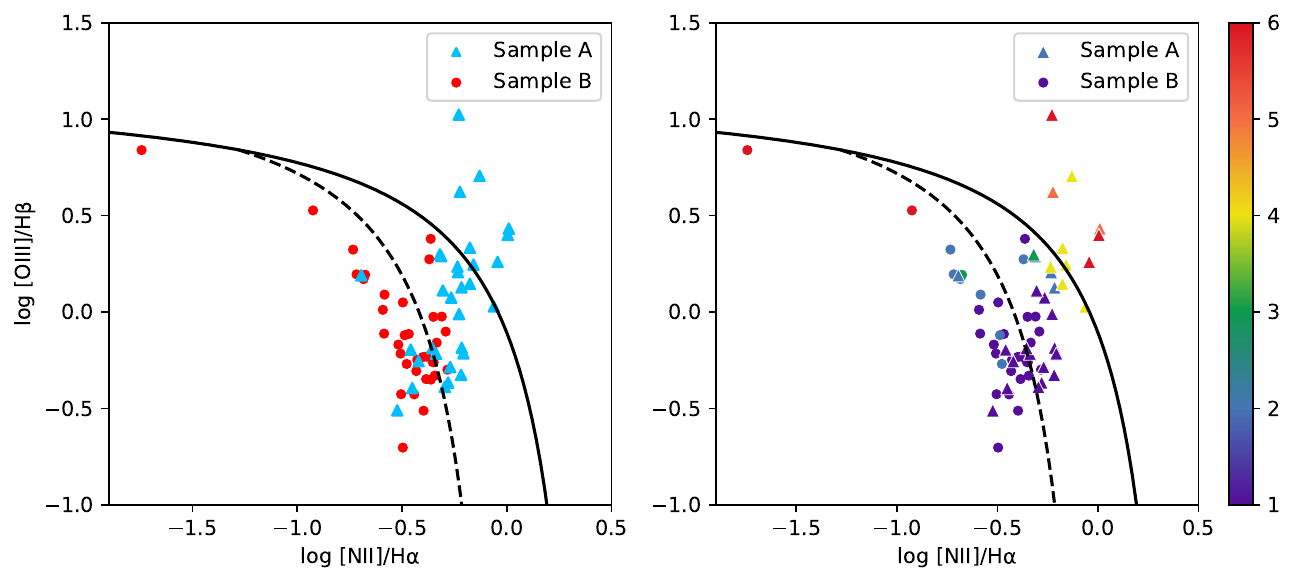}
    \caption{\textit{Left}: BPT diagram based on our single-fiber spectral analysis for the 66 galaxies in Samples A (blue triangles) and B (red circles) that have available spectra from Australian Astronomical Optics Data Central with quality emission line fits. \textit{Right}: Same diagram as the left panel but with the colors corresponding to the score each galaxy received based on the SAMI IFS data (see \S\ref{sec:selection_scheme} for details on our scoring scheme).}
    \label{fig:fib1}
\end{figure*}

We first fit the [SII] doublet with two Gaussians while fixing the separation between the peaks to the laboratory value and demanding that both lines have the same width in velocity space. We then fit the [NII]+$H{\alpha}$ complex with three Gaussians. We fix the widths of the [NII] doublet lines to be equal in velocity space, we fix the separation of the doublet lines to laboratory values, and we fix the relative flux ratio of the lines to the theoretical value of 2.96. The widths of the [NII] doublet lines are required to match the width of the [SII] doublet in velocity space, whereas the width of the $H\alpha$ line is allowed to increase up to $25\%$ of the [SII] width. 

We also test for broad $H{\alpha}$ emission in the spectra by refitting the [NII]+$H{\alpha}$ complex with an additional (broader) Gaussian component. The presence of broad $H\alpha$ emission is often used as an AGN indicator, as it can originate from high velocity gas in the broad line region surrounding an accreting black hole. However, broad $H\alpha$ can also arise from other transient or energetic processes, such as supernovae \citep{baldassare_multi-epoch_2016}.
In our emission line fitting process, the broad line component is allowed to shift in velocity relative to the narrow line center to account for potential outflows, and its width is required to be at least as large as that of the narrow component. To classify a galaxy as exhibiting broad line emission, we impose three criteria: (1) the broad component must have a FWHM of at least $500 \, \text{km/s}$, (2) its flux must be detected with S/N $>$ 3, and (3) the reduced $\chi^2$ of the [NII]+$H\alpha$ model including the broad component must at least $20\%$ lower than the reduced $\chi^2$ of the model without a broad component. $H\beta$ is fit with the same approach as $H\alpha$. We do not find sufficient evidence of broad $H\alpha$ or $H\beta$ emission in any galaxy.

Finally, we fit the [OIII]$\lambda5007$ line with no constraints on the width of the line. This is because it is common for the [OIII] profile to exhibit a broad blue shoulder that does not match the other lines in our fitting process \citep{heckman_emission-line_1981, whittle_narrow_1985}. An example of spectral fitting for Sample A galaxy CATID 324517 is shown in Figure \ref{fig:spec_grid}.

After measuring all emission line fluxes for the available 76 spectra, we classify each galaxy on the [NII] diagram. If any of the four narrow emission lines required for the [NII] diagram had S/N $<$ 3, then the classification was considered unreliable. Galaxies with uncertain classifications are labeled as "Poor Fit".

Of the 41 galaxies in Sample A, the single-fiber [NII] diagram classification gives 5 SF galaxies, 18 Composite, 8 AGN, 6 galaxies with no available spectra, and 4 galaxies with poor spectral fits. Of the 46 galaxies in Sample B, there are 28 SF galaxies, 7 Composite, 0 AGN, 5 with no spectra, and 6 with poor fits based on the single-fiber spectra. The visual breakdown is shown in Figure \ref{fig:1fib_piechart}. All line fits are visually inspected to confirm the goodness of fit for the emission line measurements and to verify the absence of definite broad line features.

If we restrict our analysis to the 66 galaxies with reliable spectra and fits, we find that the vast majority of Sample B galaxies would be classified as star-forming (SF) in single-fiber diagnostic analysis. As shown in the left panel of Figure \ref{fig:fib1}, most Sample B galaxies lie along the left arm of the characteristic V-shape of the [NII] diagram, in the SF region. In contrast, 26 of the 31 fit galaxies in Sample A are in the Composite or AGN regions and seem to fall on the right arm of the diagram. Visual inspection of spatially resolved diagnostic maps reveals that many of the Sample A galaxies in the SF region exhibit irregular, off-nuclear AGN emission, which would likely be missed by single-fiber surveys. We also note that these galaxies occupy lower scoring bins (1 or 2). The higher incidence of such decentralized emission in Sample B galaxies explains why more of them appear SF in the single-fiber classification despite harboring AGN or Composite emission regions. A discussion of this apparent decentralized AGN emission is given below in \S\ref{sec:decentralized_agn}. The right panel of Figure \ref{fig:fib1} further shows that galaxies with higher AGN likelihood scores tend to lie near or above the maximum starburst line on the [NII] diagram, consistent with AGN ionization. Two high scoring exceptions from sample B fall in the low-metallicity, far-left region of the diagram, these are also discussed in \S\ref{sec:decentralized_agn}. Galaxies labeled "Poor Fit" do not appear in the [NII] diagrams in Figure \ref{fig:fib1}.

\section{Discussion} \label{sec:discussion}

\subsection{Decentralized AGN Emission} \label{sec:decentralized_agn}

Galaxies in our AGN samples display a range of emission distributions in the SAMI IFS data. In some systems the AGN/Composite emission is concentrated near the nucleus, while in others it extends to larger radii and does not peak at the galactic center. We refer to such systems as having "decentralized" AGN emission.

\begin{figure}
    \centering
    \includegraphics[width=\linewidth]{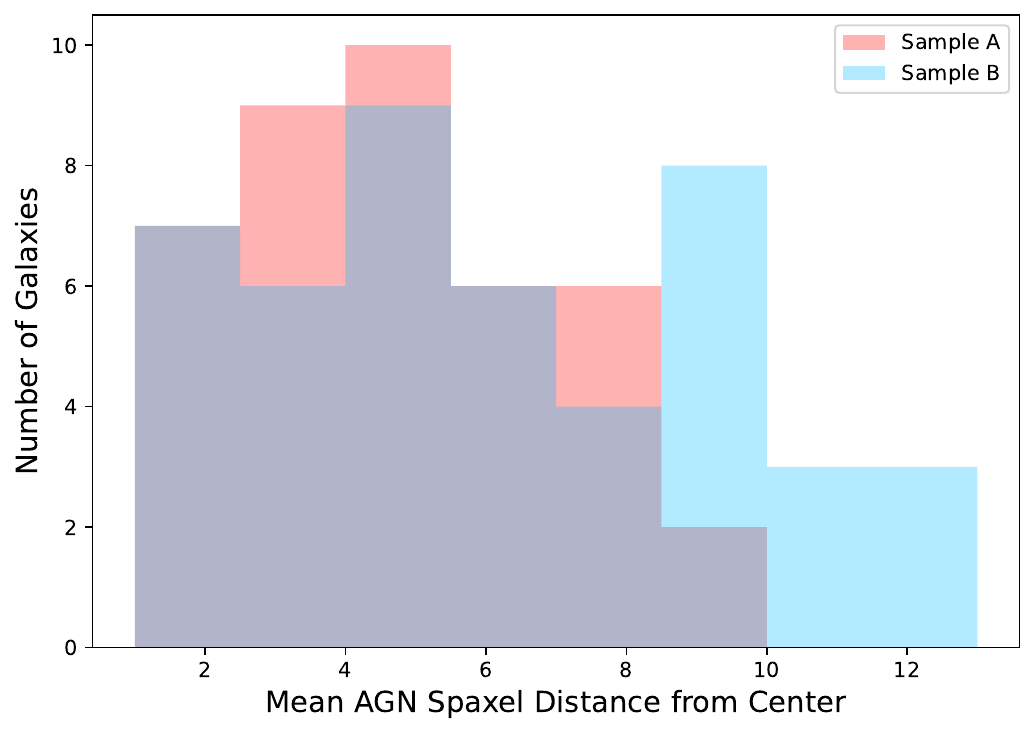}
    \caption{Distribution of the mean radial distance in spaxels of [NII]-classified AGN or Composite spaxels from the galaxy center for galaxies in Sample A (red) and Sample B (light blue).}
    \label{fig:mean_spaxel_dist}
\end{figure}

To quantify this behavior, we compute the mean radial distance of AGN/Composite spaxels from the galaxy center for each galaxy. The resulting distribution of mean radii spans a broad and continuous range in both Samples A and B (Figure \ref{fig:mean_spaxel_dist}), with no clear bimodal separation between centrally concentrated and extended systems. Although finer binning reveals a mild decrease in counts near $\sim7$ spaxels, this feature is not robust to modest changes in bin width and we therefore do not interpret it as evidence for distinct centralized and decentralized subpopulations. Galaxies with mean AGN spaxel distances above $\sim7$ spaxels frequently exhibit visually extended structures. However, some systems above this value contain both nuclear and off-nuclear emission, indicating that mean radius alone does not uniquely characterize morphology.

We also test whether decentralized emission correlates with AGN score. We find no significant relationship between AGN score and mean AGN spaxel distance; galaxies in each score bin span a comparable range of mean radii. This suggests that the scoring metric primarily reflects multi-diagnostic classification strength rather than the spatial concentration of AGN emission.

The mean AGN spaxel distances in Sample A (median 4.2 spaxels) are modestly smaller than those in Sample B (median 5.6 spaxels). However, a Kolmogorov--Smirnov test does not indicate that the two samples are drawn from statistically distinct parent distributions. Although Sample B systems tend to exhibit slightly more extended AGN/Composite emission, this difference is not significant at the level of our current sample size. We therefore find no evidence that decentralized structure is strongly linked to proximity to the demarcation boundaries in the [NII] diagnostic or to overall AGN classification confidence.

Several physical scenarios could produce decentralized AGN emission. The narrow-line region may extend well beyond the galactic center, allowing AGN-ionized gas to be detected at large radii even when nuclear emission is obscured. Alternatively, AGN-driven outflows may ionize gas in the outer regions of the galaxy, producing Composite or AGN emission away from the nucleus. Because these mechanisms can generate spatially extended emission, we do not require AGN signatures to be centrally concentrated for inclusion in our sample.

Metallicity effects may also influence the apparent spatial distribution of AGN emission. Because the [NII] diagnostic is sensitive to metallicity, AGN in low-metallicity environments can fall within the SF region of the diagram \citep{groves_emission-line_2006, izotov_broad_2007}. In such cases, AGN emission may be more clearly identified in the [SII] and [OI] diagrams, which are less sensitive to metallicity. Though we also note that decentralized emission appears in all three diagrams.

\begin{figure*}
    \centering
    \includegraphics[width=0.9\textwidth]{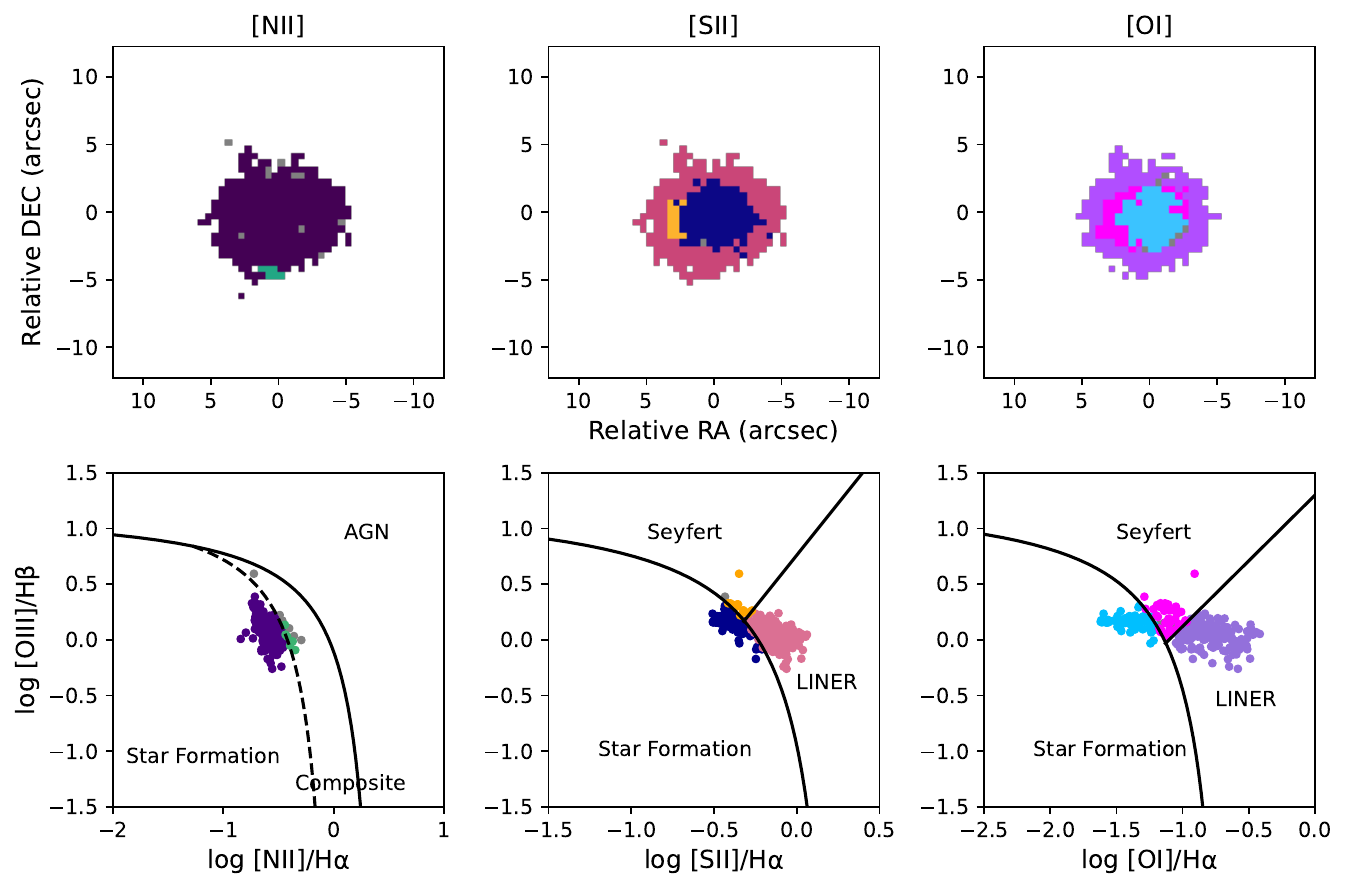}
    \caption{Example of a less secure AGN candidate in Sample B (CATID 517273). The top row shows the distribution of spaxels for each of the three diagrams and the bottom row shows the corresponding diagnostic diagrams. The color code follows Figure \ref{fig:287827_bpt}.  The galaxy is in Sample B because it displays more than 5 Composite spaxels in the [NII] diagram, but the uncertainties make it consistent with being in the SF region of the diagnostic diagram. Additionally, we are skeptical of the AGN nature of the galaxy due the Composite spaxels appearing away from the galactic center.}
    \label{fig:sparse_ex}
\end{figure*}

\begin{figure*}
\centering
\includegraphics[width=0.8\textwidth]{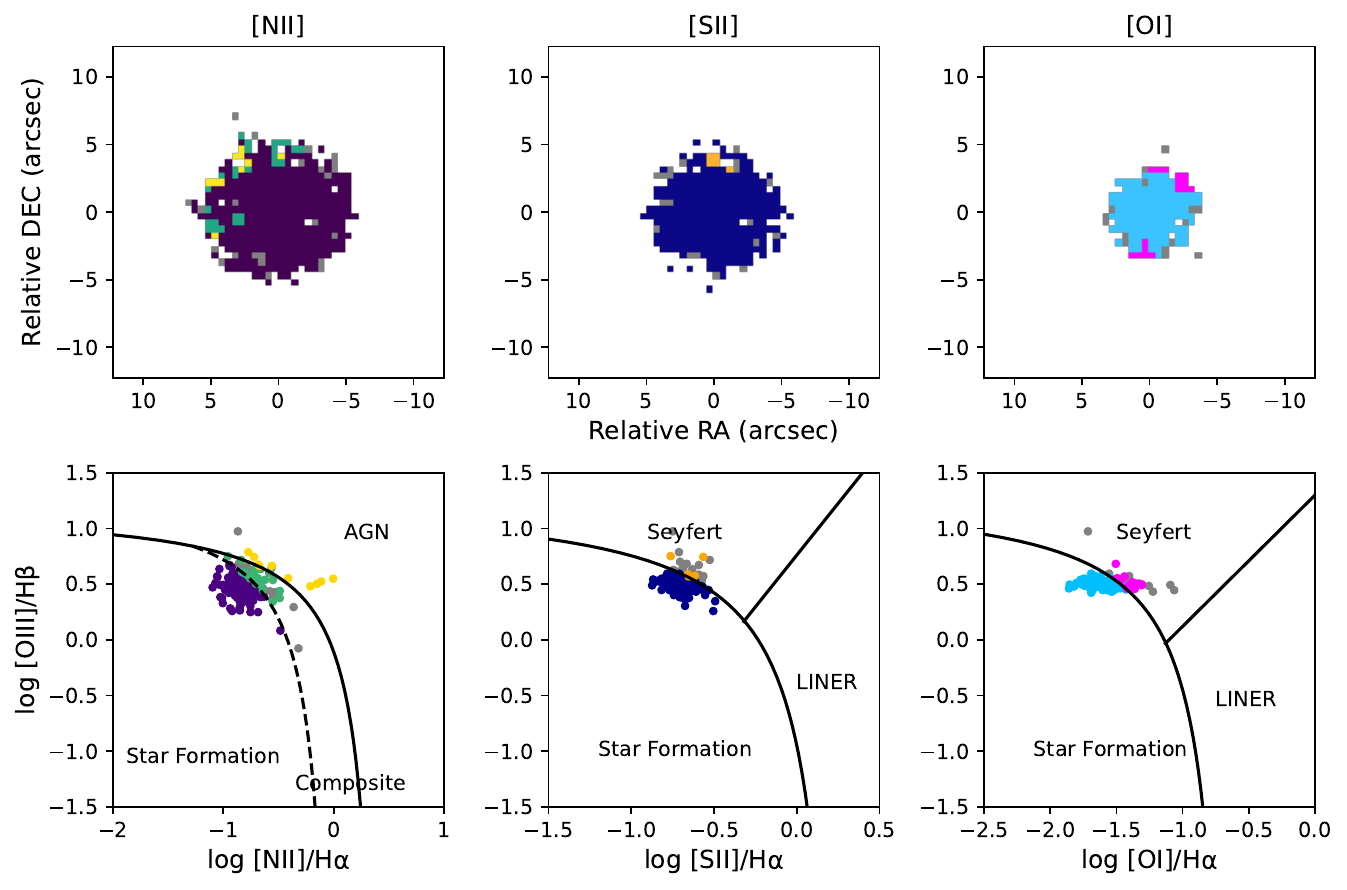}

\includegraphics[width=0.8\textwidth]{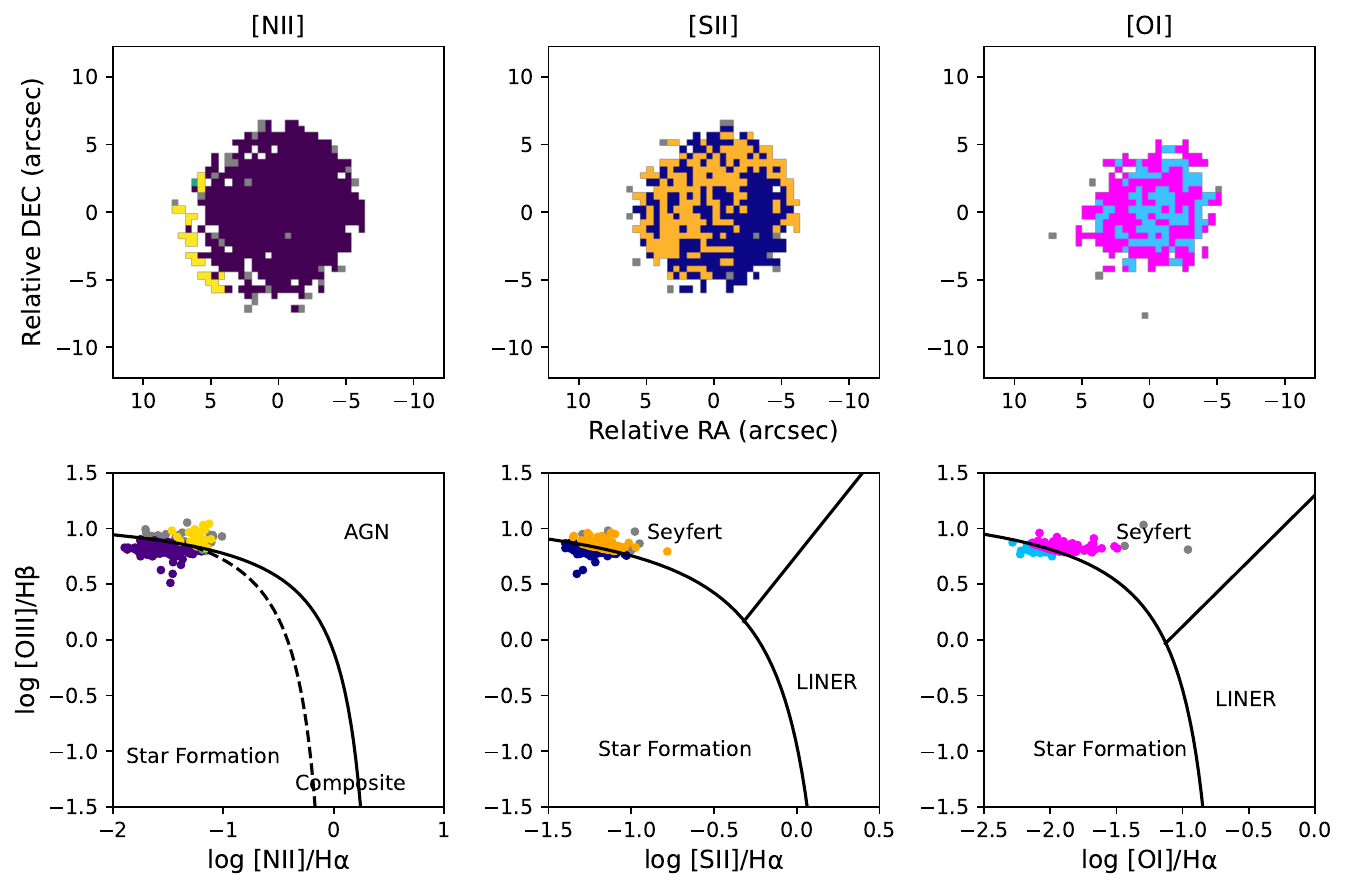}

\caption{Diagnostic maps and diagrams of two highest scoring galaxies in Sample B galaxies: SAMI CATID 40765 (top 2 rows) and CATID 493702 (bottom 2 rows). 
See Figure \ref{fig:287827_bpt} for color code explanation. 
Both galaxies display decentralized AGN emission in the [NII] diagnostic map, while CATID 40765 also shows decentralized emission in the [SII] and [OI] diagnostic maps.}
\label{fig:decentralized_bpt}
\end{figure*}

The distribution of spaxels within the diagnostic diagrams provides additional context. In several decentralized galaxies the spaxels are spread horizontally across the diagrams, particularly in the [OI] diagnostic. A similar phenomenon was discussed by \citet{johnston_beyond_2023}, who identified "AGN-like" galaxies whose spaxels form extended horizontal structures across multiple diagnostic regions. These systems were often low-mass late-type galaxies, similar to many galaxies in our Sample B. Visual inspection of our data shows that this horizontal spaxel distribution frequently occurs in galaxies with decentralized emission (e.g., Figures \ref{fig:sparse_ex} and \ref{fig:decentralized_bpt}).

Examples of decentralized emission illustrate the diversity of these systems. Many galaxies in Sample B contain predominantly SF spaxels with only a small number of Composite spaxels located away from the nucleus. An example is SAMI galaxy CATID 517273 (Figure \ref{fig:sparse_ex}), which satisfies our AGN selection criteria through the presence of sufficient Composite spaxels in the [NII] diagram and supporting classifications in the [SII] and [OI] diagrams. However, these Composite spaxels are located in the outskirts of the galaxy and are vastly outnumbered by SF spaxels, making the presence of a true AGN uncertain.

More extended AGN emission is observed in two high-scoring galaxies in Sample B: SAMI CATID 40765 and CATID 493702 (Figure \ref{fig:decentralized_bpt}). Both systems contain numerous AGN spaxels distributed well beyond the central region. CATID 493702 in particular shows AGN line ratios in the [SII] and [OI] diagrams across much of the galaxy, while CATID 40765 displays a somewhat less uniform but still extended distribution. Both galaxies occupy the upper-left region of the diagnostic diagrams and are separated from the bulk of the sample. In CATID 493702, several SF spaxels exhibit line ratios consistent with low-metallicity AGN, suggesting that the [NII] diagnostic may under represent the true extent of AGN emission in this system.

Shock ionization appears unlikely to explain the emission in these high-scoring galaxies. In both systems a substantial number of spaxels lie in the Seyfert region of the [SII] and [OI] diagrams. As shown by \citet{hong_constraining_2013}, the [SII] diagnostic is particularly effective at distinguishing between shock-ionized and photoionized gas across a wide range of metallicities. The spaxels in our galaxies fall well to the left in the [SII] diagram ($\log \text{[SII]/H}\alpha < -0.5$), beyond the limits typically associated with shock-ionized gas. If shocks are present, they are more likely driven by AGN outflows than by star formation.

In contrast, lower-scoring decentralized galaxies in either Sample A or B tend to display fewer Seyfert spaxels and more LINER classifications in the [SII] and [OI] diagrams. These systems more frequently fall within the parameter space associated with potential shock ionization, suggesting a greater likelihood of shock contributions, though such shocks could still plausibly originate from AGN-driven outflows.

Given these possibilities, we treat galaxies with decentralized emission as legitimate AGN candidates. The spatial distribution of AGN emission alone does not uniquely identify the underlying ionization mechanism, and several physically plausible processes can produce extended emission. A detailed study of these decentralized systems is beyond the scope of this work, but merits further investigation. For the purposes of this analysis, we therefore retain galaxies with decentralized emission in both Samples A and B.

\subsection{Comparison to Previous Studies} \label{sec:previous_studies}

Our results show both agreement and divergence with previous low-mass AGN studies, depending on how galaxies were classified and which diagnostics were prioritized. For galaxies where AGN emission is clearly nuclear and well-detected, our scheme yields consistent results with past single-fiber and IFS studies.

Three galaxies from the final low-mass AGN sample of \citet{salehirad_hundreds_2022}, based on GAMA DR4 spectra \citep{driver_galaxy_2022}, have SAMI data (two of which are in our AGN sample). Their AGN sample was constructed using a suite of diagnostics, including the [NII] diagram. One galaxy, CATID 323505, labeled as Composite in \citet{salehirad_hundreds_2022} also receives a Composite score of 2 in our framework. Another galaxy, CATID 324157, classified as an AGN in their sample, receives the highest score of 6 in our scheme, confirming strong AGN emission. The third galaxy, CATID 323194, is labeled as SF in the [NII] diagram in \citet{salehirad_hundreds_2022}, and is scored 0 in our scheme and therefore not included in our final sample. This object does have a detected [Fe X] line \citep{molina_sample_2021} by \citet{salehirad_hundreds_2022} and is included in their sample. We also find agreement with the one galaxy, CATID 567678, shared between our sample and the SDSS DR8 study of \citet{reines_dwarf_2013}, which relied solely on the [NII] diagnostic to identify AGN candidates. That galaxy lies in the Composite region of the [NII] diagram in \citet{reines_dwarf_2013} and likewise receives a comparable score of 3 in our system. These overlaps show that our IFS selection scheme agrees with earlier single-fiber studies when AGN emission is concentrated in the nucleus.

\begin{deluxetable}{r|c}[h!]
\tabletypesize{\footnotesize}
\tablecolumns{2} 
\tablecaption{\label{tab:BPT_breakdown} Diagnostic Diagram Result Breakdown} 
\tablehead{\colhead{Diagram(s)} & \colhead{Galaxy Count (Parent Sample $\%$)}} 
\startdata
$\text{[NII]}$ & 87 ($8.8\%$)\\
$\text{[SII]}$ & 361 ($36\%$)\\
$\text{[OI]}$ & 450 ($45\%$) \\
$\text{[NII] + [SII]}$ & 12 ($1.2\%$)\\
$\text{[NII] + [OI]}$ & 19 ($1.9\%$)\\
$\text{[SII] + [OI]}$ & 277 ($28\%$)\\
$\text{[NII] + [SII] + [OI]}$ & 8 ($0.8\%$)\\
\enddata 
\tablecomments{The total number of galaxies within the parent sample of 990 low-mass galaxies, along with the approximate percentage of the parent sample, that meet the AGN selection requirements for each diagram or combination of diagrams. Selection requirements are described in greater detail in \S\ref{sec:parent_sample} and \S\ref{sec:selection_scheme}. Briefly, at least 5 valid spaxels in the AGN or Composite regions are needed to be selected in the [NII] diagram, and at least 5 valid spaxels are required in the Seyfert region for the [SII] and [OI] diagrams. Note that we do not demand [NII] AGN or Composite detection for the [SII] and [OI] results in this table.}
\end{deluxetable}

Greater discrepancies arise when comparing with the low-mass AGN sample of \citet{mezcua_manga_2024} (hereafter MS24), which identified dwarf AGN candidates using all three optical diagnostic diagrams in the MaNGA IFS survey. Of the 28 galaxies in the final MS24 dwarf AGN sample that also appear in the SAMI survey, 11 do not satisfy the S/N requirements for classification in our analysis, 14 are classified as SF, and only 3 appear in our final AGN samples. The three overlapping galaxies are classified as AGN in MS24 and receive scores of 4, 1, and 1 in our framework.

Among the 14 galaxies classified as SF in our scheme, MS24 labels 1 as “AGN,” 2 as “Composite,” and 11 as “SF–AGN,” meaning that they show SF emission in the [NII] diagram but AGN emission in either the [SII] or [OI] diagrams. This difference reflects a key distinction between the two selection approaches. In our framework, galaxies lacking AGN or Composite emission in the [NII] diagram are classified as SF regardless of classifications in the other diagnostics. In contrast, MS24 allows galaxies with SF classifications in [NII] but AGN signatures in [SII] or [OI] to be considered AGN candidates.

This distinction has a significant impact on the inferred AGN population. As shown in Table \ref{tab:BPT_breakdown}, roughly one third to one half of galaxies in our parent sample exhibit AGN emission in the [SII] or [OI] diagrams alone. Without confirmation in the [NII] diagnostic, however, such detections are likely to include substantial contamination from non-AGN ionization processes. Differences in instrumental characteristics and S/N thresholds between SAMI and MaNGA may also contribute to smaller discrepancies. Overall, this comparison illustrates how differences in diagnostic priorities and spaxel-level selection criteria can strongly influence AGN identification in IFS surveys.

Our AGN fractions fall within the range reported in other IFS studies. Considering only galaxies in Sample A, our AGN fraction of $\sim4\%$ consistent with the $\sim5\%$ reported by \citet{wylezalek_sdss-iv_2018} and \citet{mezcua_hidden_2020}. Including Sample B as well raises our fraction to $9\%$, which remains noticeably less than $\sim20\%$ fraction (or more than $50\%$ including SF–AGN) found by MS24. These IFS AGN fractions are systematically higher than those reported in single-fiber studies, such as the $\lesssim 1-2\%$ fraction in recent work \citep{reines_dwarf_2013,salehirad_hundreds_2022,pucha_tripling_2025}, which is expected given the increased sensitivity to AGN emission on smaller spatial scales with IFS data and the ability of IFS to capture extended or decentralized AGN emission. Indeed, our comparisons show that even when classifications agree, single-fiber data miss the spatial complexity of AGN signatures revealed by IFS. Thus, while single-fiber spectra can flag strong nuclear AGN, only IFS enables robust identification of low-mass AGN candidates and the characterization of their emission distributions.

\section{Conclusions} \label{sec:conclusion}

In this work, we developed a novel automated algorithm to systematically search for AGN candidates in 990 low-mass galaxies with $M_* \leq 10^{10}M_{\odot}$ and $z \leq 0.114$. We analyzed IFS data from the SAMI Galaxy Survey (DR3) using optical emission line diagnostic diagrams. We identified 87 galaxies in the mass range $8.5 \leq \log M_*/M_{\odot} \leq 10$ and redshift range $0.006 \leq z \leq 0.058$ that show sufficient evidence of AGN or Composite emission to meet our selection criteria presented in \S\ref{sec:selection_scheme}. This results in a high confidence AGN fraction of $\sim 4\%$ ($\sim 9\%$ including low confidence galaxies) of the parent sample of 990 galaxies. Our main results are summarized in Table \ref{tab:sample_stats}.

These objects are split into two sub-samples. Sample A has 41 galaxies with high confidence AGN classifications, and Sample B has 46 galaxies with lower confidence AGN classifications. The primary diagnostic used in this paper is the $[OIII]\lambda5007/H\beta$ vs.\ $[NII]\lambda6583/H\alpha$ ([NII] BPT) diagram. The $[OIII]\lambda5007/H\beta$ vs.\ $[SII]\lambda\lambda6717,6331/H\alpha$ and $[OIII]\lambda5007/H\beta$ vs.\ $[OI]\lambda6300/H\alpha$ ([SII] and [OI] VO87) diagrams were used as supplementary diagnostics to the [NII] diagram. Our AGN candidates are assigned between 1 and 6 points in our selection scheme (\S\ref{sec:selection_scheme}), with higher scores indicating more evidence for an AGN.
While the diagnostic diagrams and selection scheme in this work provide a relatively clean sample (particularly Sample A), this analysis is susceptible to missing weak AGNs and AGNs in galaxies with strong ongoing star-formation.

For comparison, we also analyze single-fiber spectra available for Samples A and B and place them on the [NII] diagnostic diagram. We find that, of the galaxies that had a good quality single-fiber spectrum, $\sim 16\%$ of the galaxies in Sample A and $\sim 80\%$ of the galaxies in Sample B are classified as star-forming dominated. This shows the ability of IFS data to identify AGN signatures in galaxies that display a decentralized distribution of AGN emission (see \S\ref{sec:decentralized_agn}), whereas single-fiber surveys are typically restricted to the central region of the galaxy. In the cases where AGN emission is dominant and concentrated in the center of a galaxy, single-fiber spectra and IFS data produce similar results.

This paper provides a new sample of low-mass AGN candidates near the celestial equator using IFS data. IFS surveys are capable of probing AGN behavior through spatially resolved emission maps, enabling the detection of atypical AGN emission distributions. Follow-up observations (e.g., with JWST and/or Chandra) would be useful to confirm the presence of AGNs in these galaxies and better characterize the hosts. In general, further research into AGN candidates with available IFS data is warranted, as is the continued expansion of IFS survey coverage. \newline

The authors would like to thank the anonymous referee for their helpful comments.
A.E.R. gratefully acknowledges support for this work provided by NSF through CAREER award 2235277.

The SAMI Galaxy Survey is based on observations made at the Anglo-Australian Telescope. The Sydney-AAO Multi-object Integral field spectrograph (SAMI) was developed jointly by the University of Sydney and the Australian Astronomical Observatory. The SAMI input catalogue is based on data taken from the Sloan Digital Sky Survey, the GAMA Survey and the VST ATLAS Survey. The SAMI Galaxy Survey is funded by the Australian Research Council Centre of Excellence for All-sky Astrophysics (CAASTRO), through project number CE110001020, and other participating institutions. The SAMI Galaxy Survey website is http://sami-survey.org/.

GAMA is a joint European-Australasian project based around a spectroscopic campaign using the Anglo-Australian Telescope. The GAMA input catalog is based on data taken from the Sloan Digital Sky Survey and the UKIRT Infrared Deep Sky Survey. Complementary imaging of the GAMA regions is being obtained by a number of independent survey programmes including GALEX MIS, VST KiDS, VISTA VIKING, WISE, Herschel-ATLAS, GMRT and ASKAP providing UV to radio coverage. GAMA is funded by the STFC (UK), the ARC (Australia), the AAO, and the participating institutions. The GAMA website is http://www.gama-survey.org/. Based on observations made with ESO Telescopes at the La Silla Paranal Observatory under program ID 179.A-2004. Based on observations made with ESO Telescopes at the La Silla Paranal Observatory under program ID 177.A-3016.

Funding for SDSS-III has been provided by the Alfred P. Sloan Foundation, the Participating Institutions, the National Science Foundation, and the U.S. Department of Energy Office of Science. The SDSS-III web site is http://www.sdss3.org/.

SDSS-III is managed by the Astrophysical Research Consortium for the Participating Institutions of the SDSS-III Collaboration including the University of Arizona, the Brazilian Participation Group, Brookhaven National Laboratory, Carnegie Mellon University, University of Florida, the French Participation Group, the German Participation Group, Harvard University, the Instituto de Astrofisica de Canarias, the Michigan State/Notre Dame/JINA Participation Group, Johns Hopkins University, Lawrence Berkeley National Laboratory, Max Planck Institute for Astrophysics, Max Planck Institute for Extraterrestrial Physics, New Mexico State University, New York University, Ohio State University, Pennsylvania State University, University of Portsmouth, Princeton University, the Spanish Participation Group, University of Tokyo, University of Utah, Vanderbilt University, University of Virginia, University of Washington, and Yale University.

The DESI Legacy Imaging Surveys consist of three individual and complementary projects: the Dark Energy Camera Legacy Survey (DECaLS), the Beijing-Arizona Sky Survey (BASS), and the Mayall z-band Legacy Survey (MzLS). DECaLS, BASS and MzLS together include data obtained, respectively, at the Blanco telescope, Cerro Tololo Inter-American Observatory, NSF’s NOIRLab; the Bok telescope, Steward Observatory, University of Arizona; and the Mayall telescope, Kitt Peak National Observatory, NOIRLab. NOIRLab is operated by the Association of Universities for Research in Astronomy (AURA) under a cooperative agreement with the National Science Foundation. Pipeline processing and analyses of the data were supported by NOIRLab and the Lawrence Berkeley National Laboratory (LBNL). Legacy Surveys also uses data products from the Near-Earth Object Wide-field Infrared Survey Explorer (NEOWISE), a project of the Jet Propulsion Laboratory/California Institute of Technology, funded by the National Aeronautics and Space Administration. Legacy Surveys was supported by: the Director, Office of Science, Office of High Energy Physics of the U.S. Department of Energy; the National Energy Research Scientific Computing Center, a DOE Office of Science User Facility; the U.S. National Science Foundation, Division of Astronomical Sciences; the National Astronomical Observatories of China, the Chinese Academy of Sciences and the Chinese National Natural Science Foundation. LBNL is managed by the Regents of the University of California under contract to the U.S. Department of Energy. The complete acknowledgments can be found at https://www.legacysurvey.org/acknowledgment/.

\bibliography{ztex}
\bibliographystyle{aasjournal}

\newpage

\appendix

\section{Data Tables} \label{sec:app_tables}

We provide abbreviated versions of the summaries of Samples A and B in Tables \ref{tab:app_table} and \ref{tab:app_table_b}, respectively. The full versions of these tables are available in machine-readable format in the online version.

\begin{deluxetable}{r|cccccccccc}[h!]
\tabletypesize{\footnotesize}
\tablecolumns{2} 
\tablecaption{\label{tab:app_table}Sample A Galaxies} 
\tablehead{\colhead{CATID} & \colhead{\hspace{.75cm}RA (deg)}\hspace{.75cm} & \colhead{\hspace{.75cm}Dec. (deg)}\hspace{.75cm} & \colhead{\hspace{.33cm}$\log M_*/M_{\odot}$}\hspace{.33cm} & \colhead{z} & \colhead{Score} & \colhead{[NII]} & \colhead{[SII]} & \colhead{[OI]} & \colhead{Morph.} & \colhead{BPT Class}} 
\startdata
65384 & 222.84263733 & -0.37558728 & 9.97 & 0.04343 & 2 & 1 & 0 & 1 & EaS & COMP\\
77452 & 213.37499518 & 0.18584768 & 9.77 & 0.02584 & 1 & 1 & 0 & 0 & LaS & SF\\
79810 & 223.26107295 & 0.20313785 & 10.0 & 0.04320 & 4 & 4 & 0 & 0 & E & ---\\
91568 & 212.75914697 & 0.60849331 & 9.95 & 0.02663 & 5 & 4 & 1 & 0 & E/S0 & ---\\
92770 & 217.56241698 & 0.62137578 & 9.93 & 0.02625 & 2 & 1 & 0 & 1 & Uncertain & COMP\\
\enddata 
\tablecomments{First 5 entries of the 41 galaxies included in sample A. From left to right: SAMI catalog ID (CATID), Object Right Ascension in degrees (RA), Object Declination in degrees (Dec.), Galaxy stellar mass in solar masses on a logarithmic scale ($\log M_*/M_{\odot}$), spectroscopic redshift (z), AGN score given in this paper (Score), number of points contributing to the total score from each diagnostic diagram ([NII], [SII], [OI]), morphology of the galaxy (Morph.), single-fiber [NII] diagram classification (BPT Class). The complete version of this table is available in machine-readable format.}
\end{deluxetable}

\begin{deluxetable}{r|cccccccccc}[h!]
\tabletypesize{\footnotesize}
\tablecolumns{2} 
\tablecaption{\label{tab:app_table_b}Sample B Galaxies} 
\tablehead{\colhead{CATID} & \colhead{\hspace{.75cm}RA (deg)}\hspace{.75cm} & \colhead{\hspace{.75cm}Dec. (deg)}\hspace{.75cm} & \colhead{\hspace{.33cm}$\log M_*/M_{\odot}$}\hspace{.33cm} & \colhead{z} & \colhead{Score} & \colhead{[NII]} & \colhead{[SII]} & \colhead{[OI]} & \colhead{Morph.} & \colhead{BPT Class}} 
\startdata
8706 & 183.68935606 & 0.74353151 & 10.0 & 0.02079 & 1 & 1 & 0 & 0 & EaS & SF \\
9067 & 184.84232311 & 0.74546904 & 9.44 & 0.03369 & 1 & 1 & 0 & 0 & Ea/LaS & COMP\\
40197 & 180.03405742 & -0.66341702 & 9.04 & 0.02054 & 1 & 1 & 0 & 0 & S0/EaS & SF\\
40765 & 182.3985672 & -0.69958014 & 9.76 & 0.03512 & 6 & 4 & 1 & 1 & Ea/LaS & SF\\
41302 & 185.3985672 & -0.70289815 & 9.76 & 0.03994 & 1 & 1 & 0 & 0 & S0/EaS & SF\\
\enddata 
\tablecomments{First 5 entries of the 46 galaxies included in sample B. See Table \ref{tab:app_table} for column descriptions. The complete version of this table is available in machine-readable format.}
\end{deluxetable}

\section{Spaxel Spectra} \label{sec:spaxel_spectra}

\begin{figure*}[t!]
    \centering
    \includegraphics[width=\textwidth]{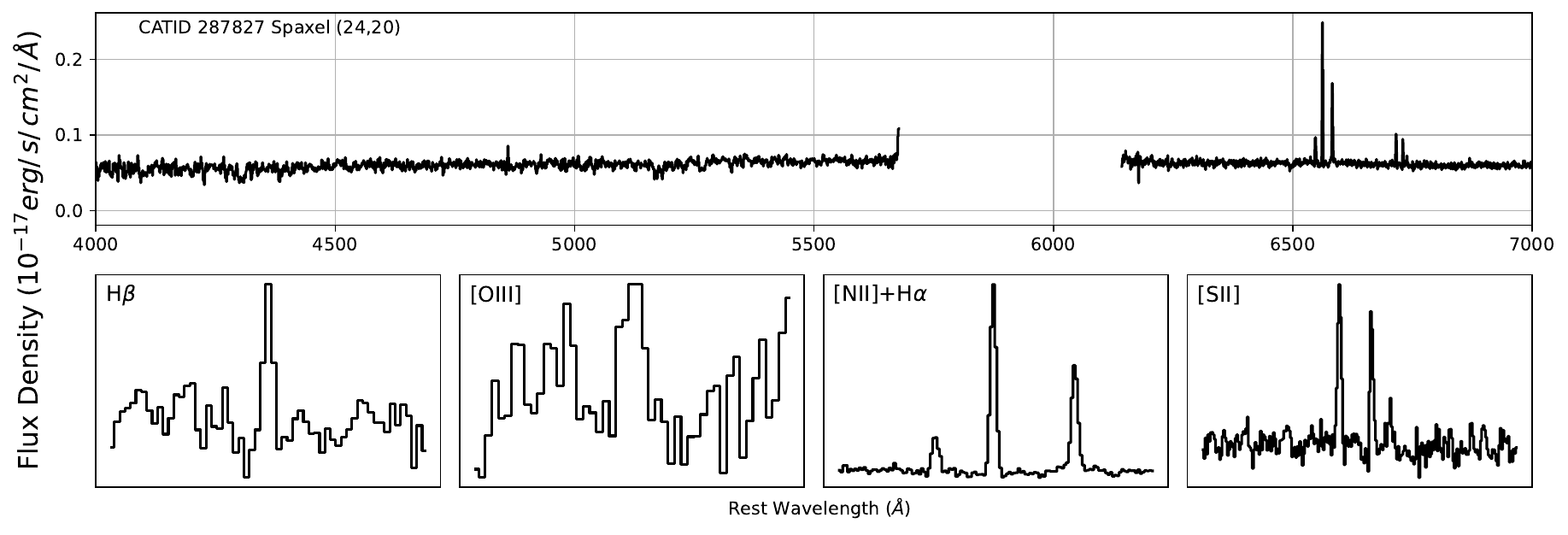}
    \caption{Top: Rest-frame blue-arm and red-arm spectra of galaxy CATID 287827 (AGN score 6) in spaxel position (24,20). Bottom: Cutouts centered on key diagnostic emission lines. Line identifications are shown in the upper left corner of each panel. These panels illustrate the emission line signal quality in a high scoring AGN spaxel.}
    \label{fig:spax_spec_287827}
\end{figure*}

\begin{figure*}[t!]
    \centering
    \includegraphics[width=\textwidth]{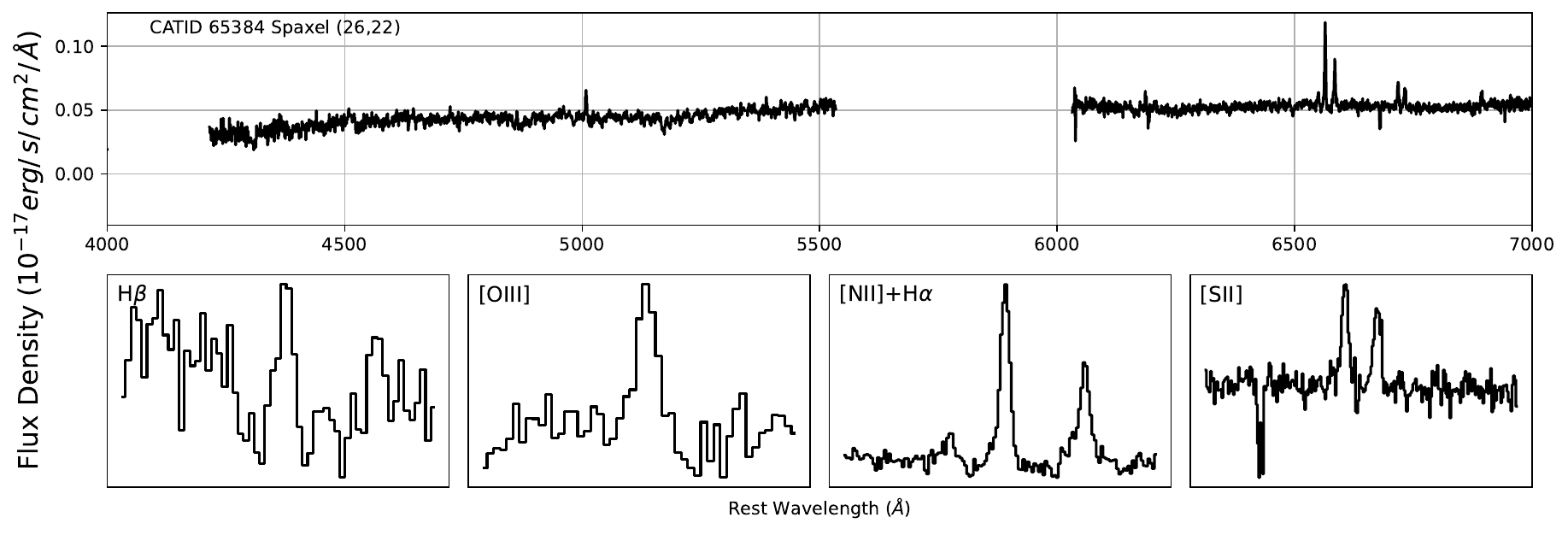}
    \caption{Same as Figure \ref{fig:spax_spec_287827}, but for galaxy CATID 65384 (AGN score 2) in spaxel position (26,22), illustrating emission line quality in a low scoring AGN spaxel.}
    \label{fig:spax_spec_65384}
\end{figure*}

Here we present spectra from randomly selected AGN spaxels in order to illustrate the quality of the spectral data for a high scoring and low scoring galaxy, shown in Figures \ref{fig:spax_spec_287827} and \ref{fig:spax_spec_65384}, respectively. Although the emission line fitting was performed as part of the SAMI data products and individual spaxel fit parameters are not available within our analysis framework, the spectra provide a direct visual assessment of emission line strength, signal-to-noise, and continuum subtraction quality. The displayed spaxels were randomly selected from spaxels from galaxies in Sample A that are classified as AGN/Composite in the [NII] diagnostic diagram.

\newpage

\section{Selection Scheme Scoring} \label{scheme_discussion}

To assess the sensitivity of our AGN scoring metric to the adopted weighting scheme, we vary the relative weights assigned to AGN emission in the three optical diagnostic diagrams. The identification of AGN candidates is determined by the requirement of sufficient AGN or Composite emission in the [NII] diagnostic. Therefore, the weighting scheme only affects the relative confidence ranking within the AGN candidate sample rather than membership of the sample.

We test alternative scoring configurations in which the [SII] or [OI] diagrams dominate, as well as a scheme in which all three diagnostics are equally weighted. Across all tested configurations, galaxies occupying the highest and lowest scoring bins (i.e., scores of 6 and 1, respectively) are largely preserved. This robustness is expected as the highest score corresponds to consistent AGN classification across all three diagnostics, while the lowest score reflects systems supported only by [NII] diagram Composite emission. The primary effect of varying the weighting scheme is therefore to redistribute galaxies among the intermediate score bins, without substantially altering the high confidence AGN population or the overall candidate sample.

Although several weighting prescriptions produce similar hierarchical samples, we adopt the [NII] dominated scheme for consistency with the broader literature. The [NII] diagram is the most widely used optical AGN diagnostic, particularly in studies of low-mass galaxies. Anchoring our scoring system to this diagram therefore facilitates direct comparison with previous work, while the [SII] and [OI] diagrams provide complementary, metallicity-sensitive probes that refine the confidence ranking. The robustness of our results under alternative weighting schemes indicates that our conclusions are not driven by this choice.

\section{Diagnostic Diagram Maps} \label{sec:bpt_maps}

Diagnostic maps and DESI legacy survey DR9 grz cutout images (DR10 for galaxy cluster objects, which have 10 digit CATIDs) are supplied for each galaxy in the sample below. The red circle in both the diagnostic maps and the cutouts represents the aperture size and location of the associated single-fiber observation for that galaxy, if available. In the [NII] diagnostic maps, the spaxels dominated by SF, Composite, and AGN emission are colored purple, green, and yellow, respectively. For [SII] maps, the SF, LINER, and Seyfert spaxels are colored blue, pale red, and orange, respectively. For [OI] maps, the SF, LINER, and Seyfert spaxels are colored light blue, purple, and pink, respectively. Gray spaxels represent spurious AGN spaxels (see \S\ref{sec:selection_scheme}, while white spaxels indicate either a lack of data or high uncertainty in the flux measurements for that spaxel. The SAMI CATID for each galaxy is printed on all diagnostic maps and the cutout. \\

\begin{figure}
    \centering
    \includegraphics[width=\textwidth]{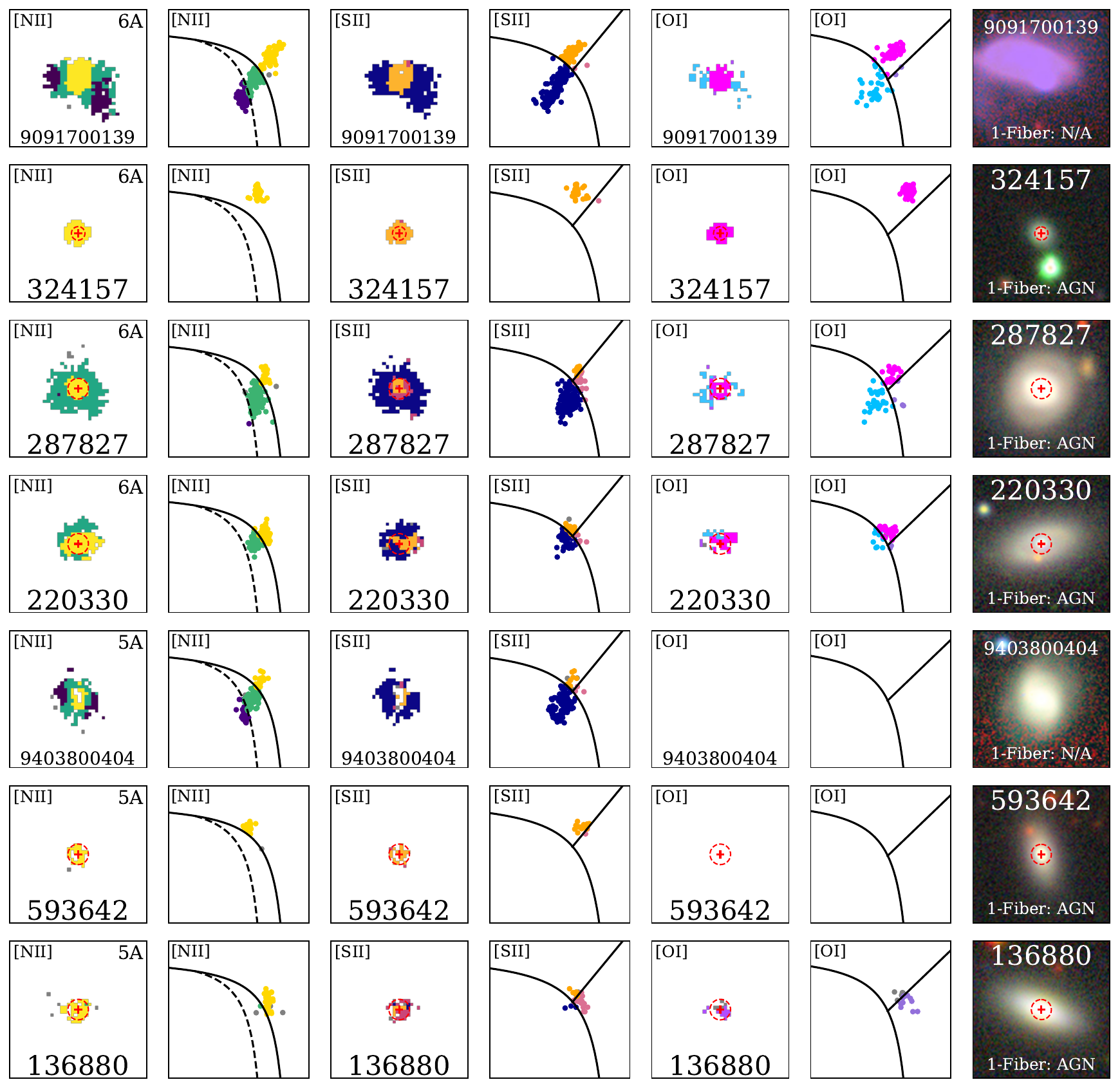}
    \caption{See Appendix \ref{sec:bpt_maps} description for details.}
    \label{fig:mapA1}
\end{figure}

\begin{figure}
    \centering
    \includegraphics[width=\textwidth]{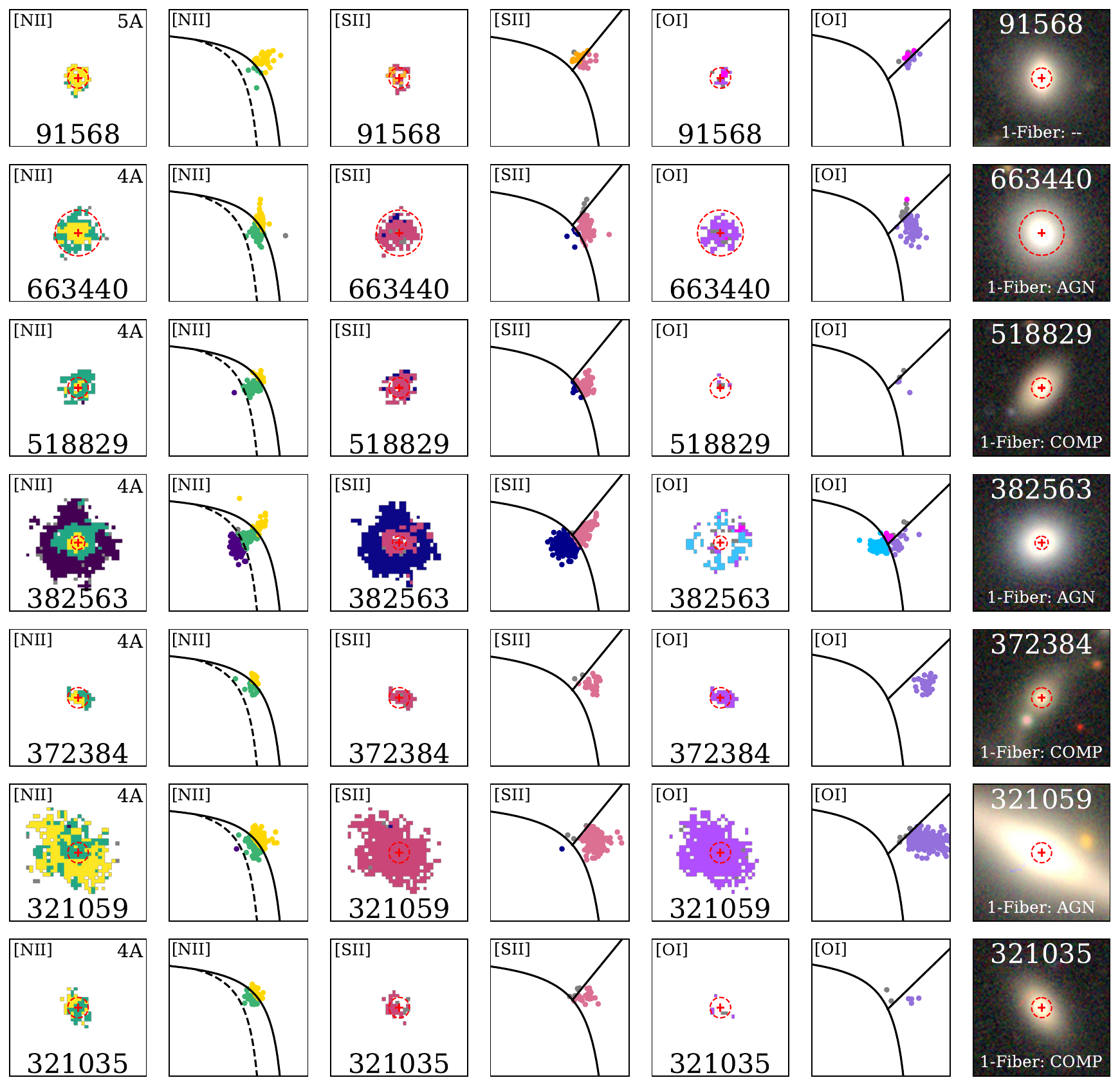}
    \caption{See Appendix \ref{sec:bpt_maps} description for details.}
    \label{fig:mapA2}
\end{figure}

\begin{figure}
    \centering
    \includegraphics[width=\textwidth]{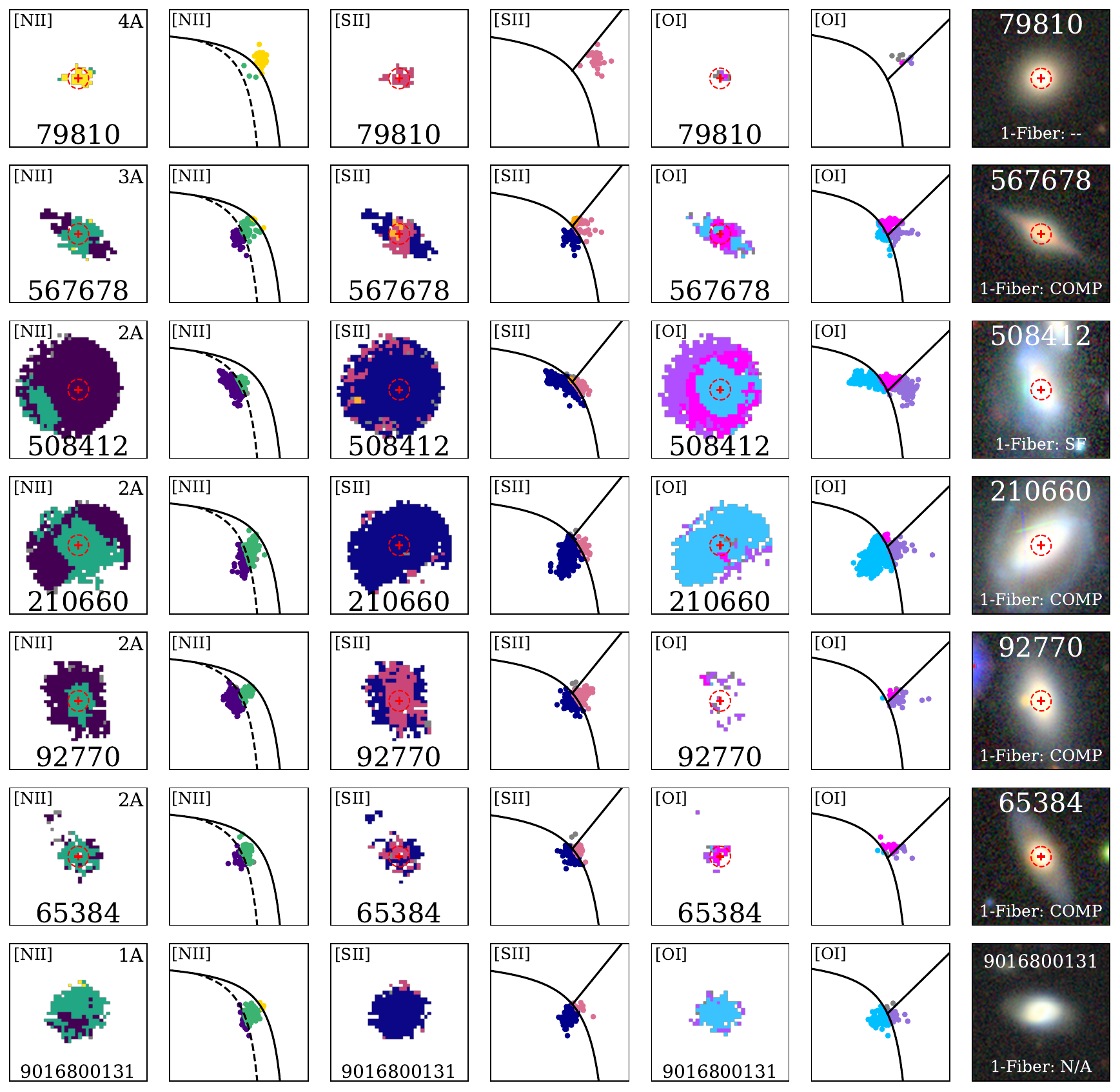}
    \caption{See Appendix \ref{sec:bpt_maps} description for details.}
    \label{fig:mapA3}
\end{figure}

\begin{figure}
    \centering
    \includegraphics[width=\textwidth]{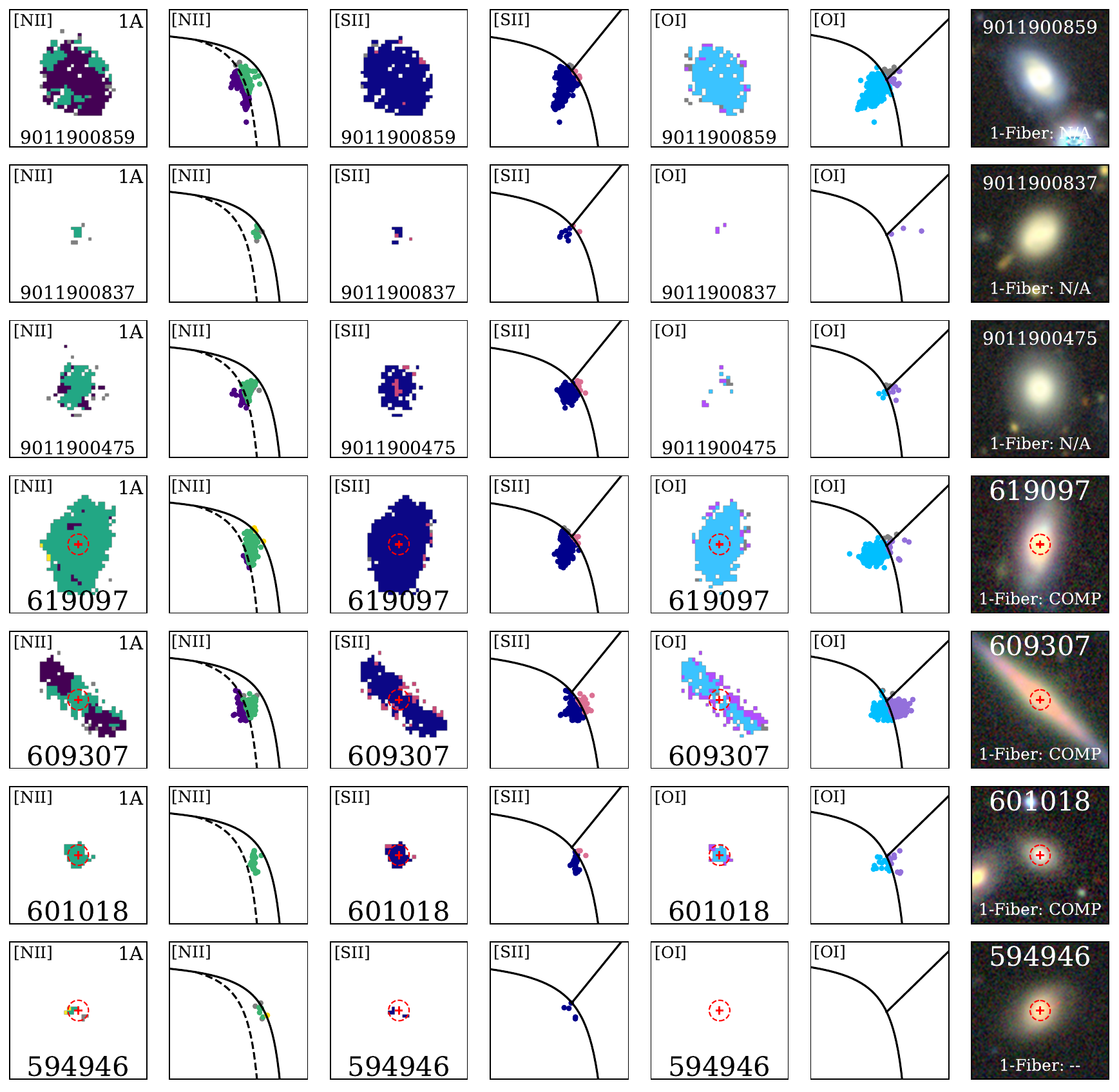}
    \caption{See Appendix \ref{sec:bpt_maps} description for details.}
    \label{fig:mapA4}
\end{figure}

\begin{figure}
    \centering
    \includegraphics[width=\textwidth]{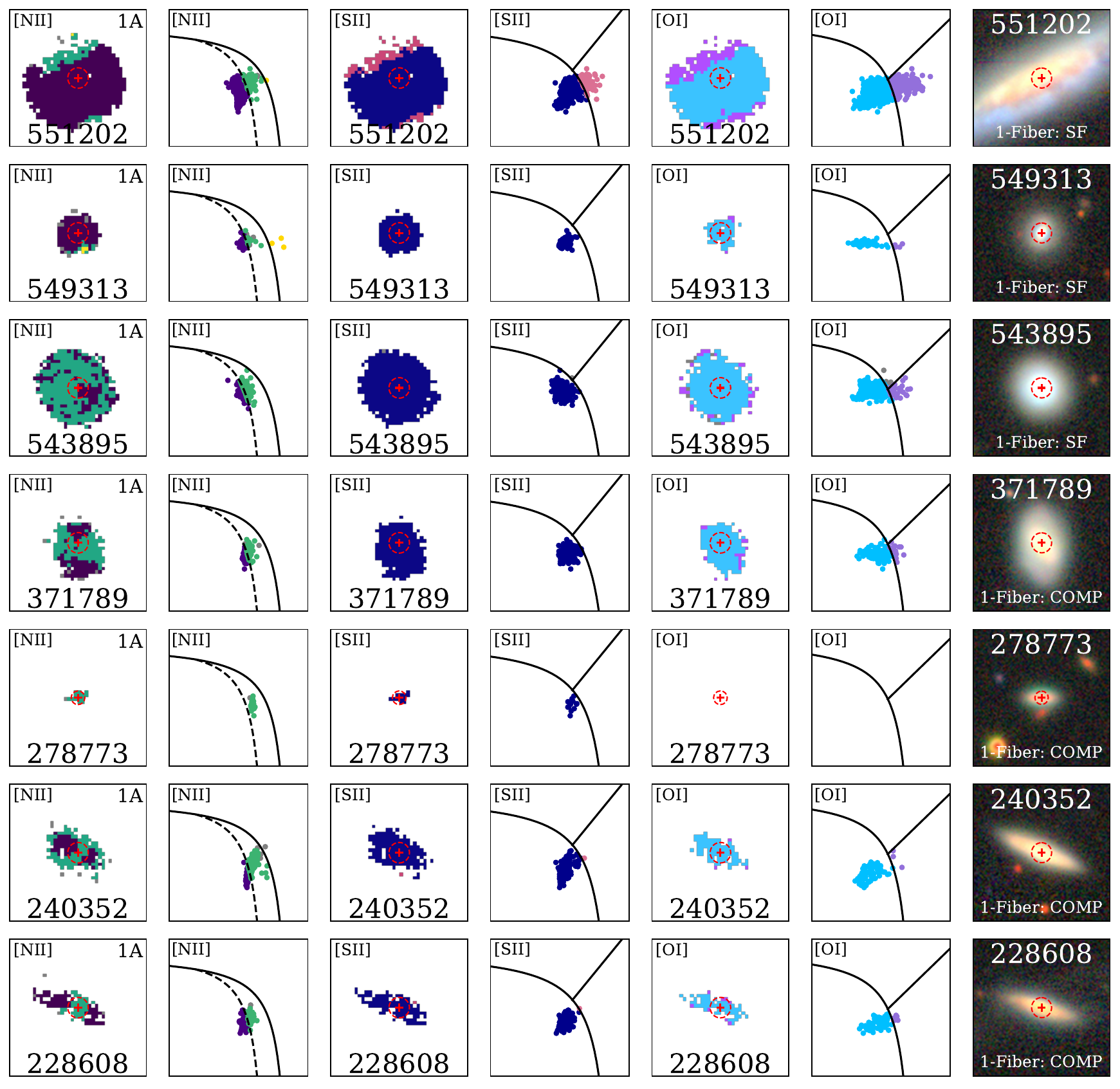}
    \caption{See Appendix \ref{sec:bpt_maps} description for details.}
    \label{fig:mapA5}
\end{figure}

\begin{figure}
    \centering
    \includegraphics[width=\textwidth]{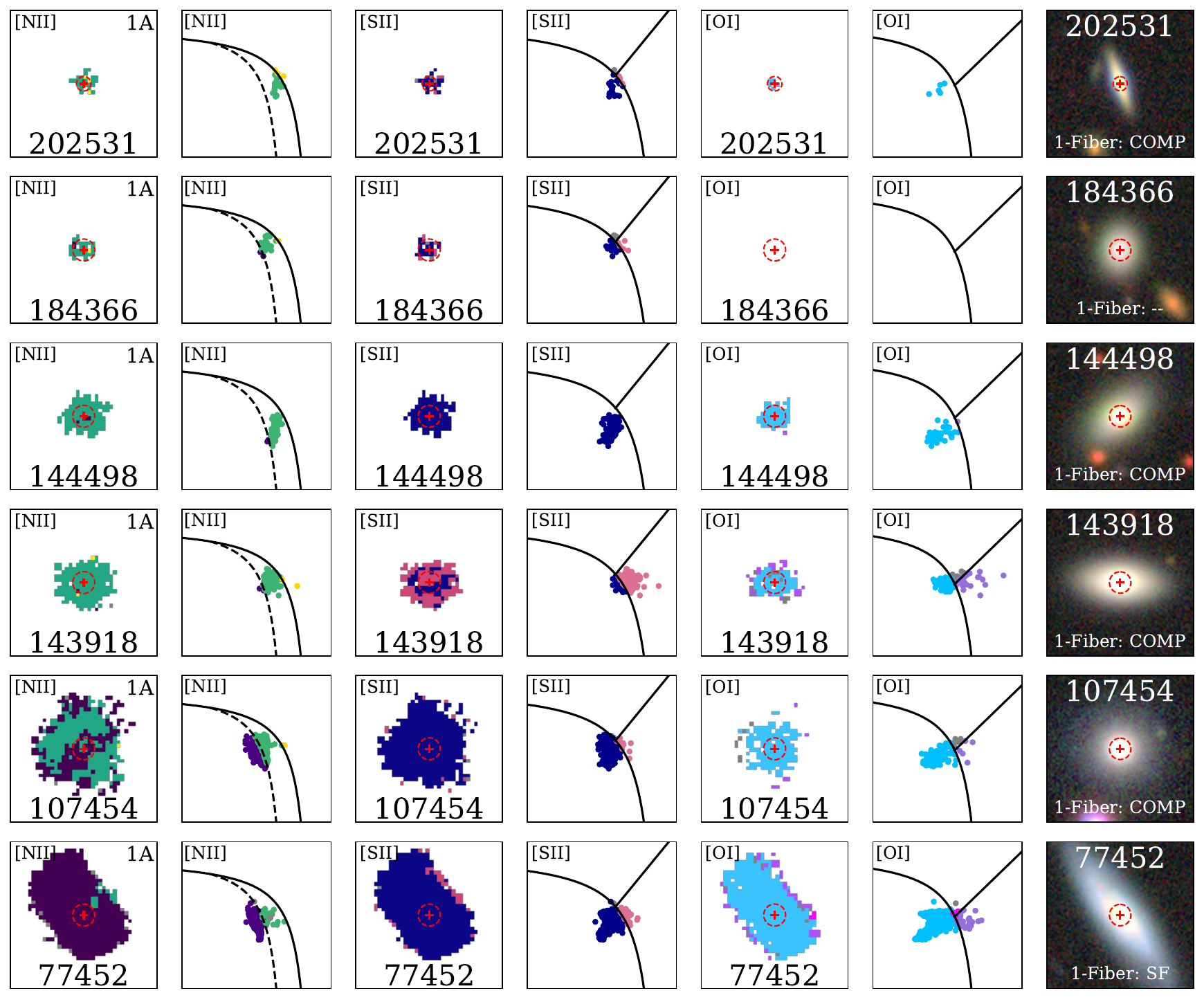}
    \caption{See Appendix \ref{sec:bpt_maps} description for details.}
    \label{fig:mapA6}
\end{figure}

The AGN score for the galaxy is given in the upper right corner of the [NII] diagnostic map. The scores range from 1 to 6, where 6 is very likely to be an AGN and 1 is less likely. All galaxies shown are in sample A, see \S\ref{sec:selection_scheme} for details. At the bottom of the DESI cutout we report the result of the single-fiber [NII]-BPT classification for that galaxy. In the cases where the uncertainties in the flux values were too large to confidently place the galaxy on the diagram, the classification is "--". Galaxies with no available single-fiber spectra are labeled "N/A". The maps are organized primarily by highest to lowest score, and then by descending CATID. Diagnostic maps and optical cutouts are presented in Figures \ref{fig:mapA1} -- \ref{fig:mapA6}.

\end{document}